\documentclass[twoside,twocolumn,9pt]{article}
\usepackage{extsizes}
\usepackage[super,sort&compress,comma]{natbib} \usepackage[export]{adjustbox}

\usepackage[version=3]{mhchem}
\usepackage[left=1.5cm, right=1.5cm, top=1.785cm, bottom=2.0cm]{geometry}
\usepackage{balance}
\usepackage{amsmath}
\usepackage{sectsty}
\usepackage{txfonts}
\usepackage{graphicx} 
\usepackage[format=plain,justification=justified,singlelinecheck=false,font={stretch=1.125,small,sf},labelfont=bf,labelsep=space]{caption}
\usepackage{float}
\usepackage{fancyhdr}
\usepackage{fnpos}
\usepackage[english]{babel}
\usepackage{array}
\usepackage{droidsans}
\usepackage{charter}
\usepackage[T1]{fontenc}
\usepackage[usenames,dvipsnames]{xcolor}
\usepackage{setspace}
\usepackage[compact]{titlesec}
\usepackage{float}
\usepackage{chemformula}
\usepackage{newtxmath}

\usepackage[resetlabels]{multibib}
\newcites{S}{References}

\def\rmlj {\rlap{\kern 0.0em\raise 1.3ex\hbox
  {\vrule height 0.10ex width 0.28em}}l}
\def\sllj {\rlap{\kern 0.25em\raise 1.3ex\hbox
  {\vrule height 0.10ex width 0.28em}}l}
\def\itlj {\rlap{\kern 0.25em\raise 1.3ex\hbox
  {\vrule height 0.10ex width 0.28em}}l}
\def\bflj {\rlap{\kern 0.25em\raise 1.3ex\hbox
  {\vrule height 0.10ex width 0.28em}}l}
\def\ttlj {\rlap{\kern 0.25em\raise 1.3ex\hbox
  {\vrule height 0.10ex width 0.28em}}l}
\def\sclj {\rlap{\kern 0.25em\raise 1.3ex\hbox
  {\vrule height 0.10ex width 0.28em}}l}
\def\sflj {\rlap{\kern 0.25em\raise 1.3ex\hbox
  {\vrule height 0.10ex width 0.28em}}l}

\def\lj{\ifvmode\leavevmode\fi\ifcase\fam \rmlj \or \or \or
  \or \itlj \or \sllj \or \bflj \or \ttlj \or \sflj \or \sclj \else \rmlj \fi}

\def\rmLj {\rlap{\kern 0.03em\raise 1.2ex\hbox
  {\vrule height 0.10ex width 0.28em}}L}
\def\slLj {\rlap{\kern 0.25em\raise 1.3ex\hbox
  {\vrule height 0.10ex width 0.28em}}L}
\def\itLj {\rlap{\kern 0.25em\raise 1.3ex\hbox
  {\vrule height 0.10ex width 0.28em}}L}
\def\bfLj {\rlap{\kern 0.25em\raise 1.3ex\hbox
  {\vrule height 0.10ex width 0.28em}}L}
\def\ttLj {\rlap{\kern 0.25em\raise 1.3ex\hbox
  {\vrule height 0.10ex width 0.28em}}L}
\def\scLj {\rlap{\kern 0.25em\raise 1.3ex\hbox
  {\vrule height 0.10ex width 0.28em}}L}
\def\sfLj {\rlap{\kern 0.25em\raise 1.3ex\hbox
  {\vrule height 0.10ex width 0.28em}}L}

\def\Lj{\ifvmode\leavevmode\fi\ifcase\fam \rmLj \or \or \or
  \or \itLj \or \slLj \or \bfLj \or \ttLj \or \sfLj \or \scLj \else \rmLj \fi}

\def\rmdj {\rlap{\kern 0.25em\raise 1.2ex\hbox
  {\vrule height 0.10ex width 0.28em}}d}
\def\sldj {\rlap{\kern 0.25em\raise 1.3ex\hbox
  {\vrule height 0.10ex width 0.28em}}d}
\def\itdj {\rlap{\kern 0.25em\raise 1.3ex\hbox
  {\vrule height 0.10ex width 0.28em}}d}
\def\bfdj {\rlap{\kern 0.25em\raise 1.3ex\hbox
  {\vrule height 0.10ex width 0.28em}}d}
\def\ttdj {\rlap{\kern 0.25em\raise 1.3ex\hbox
  {\vrule height 0.10ex width 0.28em}}d}
\def\scdj {\rlap{\kern 0.25em\raise 1.3ex\hbox
  {\vrule height 0.10ex width 0.28em}}d}
\def\sfdj {\rlap{\kern 0.25em\raise 1.3ex\hbox
  {\vrule height 0.10ex width 0.28em}}d}

\def\dj{\ifvmode\leavevmode\fi\ifcase\fam \rmdj \or \or \or
  \or \itdj \or \sldj \or \bfdj \or \ttdj \or \sfdj \or \scdj \else \rmdj \fi}

\def\saltino{\vskip.1cm}

\def\o{\overline}
\def\f{\frac}
\def\p{\partial}
\def\be{\begin{equation}}
\def\ee{\end{equation}}

\usepackage{epstopdf}

\definecolor{cream}{RGB}{222,217,201}

\begin{document}

\pagestyle{fancy}
\thispagestyle{plain}
\fancypagestyle{plain}{

\renewcommand{\headrulewidth}{0pt}
}

\makeFNbottom
\makeatletter
\renewcommand\LARGE{\@setfontsize\LARGE{15pt}{17}}
\renewcommand\Large{\@setfontsize\Large{12pt}{14}}
\renewcommand\large{\@setfontsize\large{10pt}{12}}
\renewcommand\footnotesize{\@setfontsize\footnotesize{7pt}{10}}
\makeatother

\renewcommand{\thefootnote}{\fnsymbol{footnote}}
\renewcommand\footnoterule{\vspace*{1pt}%
\color{cream}\hrule width 3.5in height 0.4pt \color{black}\vspace*{5pt}} 
\setcounter{secnumdepth}{5}

\makeatletter 
\renewcommand\@biblabel[1]{#1}            
\renewcommand\@makefntext[1]%
{\noindent\makebox[0pt][r]{\@thefnmark\,}#1}
\makeatother 
\renewcommand{\figurename}{\small{Fig.}~}
\sectionfont{\sffamily\Large}
\subsectionfont{\normalsize}
\subsubsectionfont{\bf}
\setstretch{1.125} 
\setlength{\skip\footins}{0.8cm}
\setlength{\footnotesep}{0.25cm}
\setlength{\jot}{10pt}
\titlespacing*{\section}{0pt}{4pt}{4pt}
\titlespacing*{\subsection}{0pt}{15pt}{1pt}

\fancyfoot{}
\renewcommand{\headrulewidth}{0pt} 
\renewcommand{\footrulewidth}{0pt}
\setlength{\arrayrulewidth}{1pt}
\setlength{\columnsep}{6.5mm}
\setlength\bibsep{1pt}

\makeatletter 
\newlength{\figrulesep} 
\setlength{\figrulesep}{0.5\textfloatsep} 

\newcommand{\topfigrule}{\vspace*{-1pt}%
\noindent{\color{cream}\rule[-\figrulesep]{\columnwidth}{1.5pt}} }

\newcommand{\botfigrule}{\vspace*{-2pt}%
\noindent{\color{cream}\rule[\figrulesep]{\columnwidth}{1.5pt}} }

\newcommand{\dblfigrule}{\vspace*{-1pt}%
\noindent{\color{cream}\rule[-\figrulesep]{\textwidth}{1.5pt}} }

\makeatother

\twocolumn[
  \begin{@twocolumnfalse}
\vspace{3cm}
\sffamily
\begin{tabular}{m{\linewidth} p{\linewidth} }

\noindent\Large{\textbf{Shape Plasmonics and Geometric Eigenvalues: The Crystal Field Plasmon Splitting in a Sphere-to-Cube Continuous Transition.}} \\

\\

\noindent\large{
Stefano Antonio Mezzasalma,$^{\dag}$ Marek Grzelczak,$^{\circ \; \ast}$ Jordi Sancho-Parramon$^{\dag \ast}$
                    } \\

\noindent\normalsize{

{\rm A smooth sphere-to-cube transition is experimentally, computationally and theoretically studied in plasmonic Au nanoparticles, including retardation effects.
Localized surface plasmon-polariton resonances were described with precision, discriminating among the influences of shape statistics, particle polydispersity, electrochemistry of excess (surface) charges.
Sphere, cube and semicubes in between all show well-defined secular electrostatic eigenvalues, producing a wealthy of topological modes afterwards quenched by charge relaxation processes.
The way both eigenvalues and plasmon wavelength vary as a function of a shape descriptor, parametrizing the transition, is explained by a minimal model based on the key concepts of crystal (or ligand) field theory (CFT), bringing for the first time to an {\em electromagnetic analog of crystal field splitting}.
For any orbital angular momentum, eigenvalues evolve as in a Tanabe-Sugano correlation diagram, relying on the symmetry set by particle topology and a charge defect between cube and sphere.
Expressions for non-retarded and retarded plasmon wavelengths are given and succeffully applied to both experimental UV-Vis and numerically simulated values.
The CFT analogy can be promising to delve into the role of shape in nanoplasmonics and nanophotonics.}
}

\saltino

\end{tabular}

 \end{@twocolumnfalse} \vspace{0.6cm}

 ]


\renewcommand*\rmdefault{bch}\normalfont\upshape
\rmfamily
\section*{}
\vspace{-1cm}


\footnotetext{\textit{$^{\dag}$~Materials Physics Division, Ru\dj er Bo\v{s}kovi\'c Institute, Bijeni\v{c}ka cesta 54, 10000 Zagreb, Croatia; E-mail: Stefano.Mezzasalma@irb.hr; Jordi.Sancho.Parramon@irb.hr}}
\footnotetext{\textit{$^{\ddag}$~CIC biomaGUNE, Miramon Pasealekua 182, 20014 Donostia-San Sebasti\'{a}n, Spain; E-mail: asanchez@cicbiomagune.es;} \textit{$^{\circ}$~Donostia International Physics Center (DIPC), Manuel Lardizabal Ibilbidea 4, 20018 Donostia-San Sebasti\'{a}n, Spain; E-mail: marek.grzelczak@dipc.org}; \textit{$^{\ast}$~Corresponding Authors.}}








Propagating and localized surface plasmon-polaritons (LSP) are coherent collective excitations by which photons couple to quasi-free metal electrons.
As light may be confined to a smaller scale than the photon wavelength, photonic and electronic characteristic lengths can be tuned into a nanoscale device.\cite{at10,barnes03,liz-marzan_growing_2017}
Plasmonics arises from a complex interplay of electronic and geometric properties, where size and shape,\cite{grzelczak_shape_2008} chemical composition,\cite{myroshnychenko_modelling_2008} surface charge \cite{novo_direct_2008} and dielectric environment\cite{kelly_optical_2003} turn out to dramatically affect LSP resonance and absorption efficiency in plasmon-induced phenomena such as hot carriers injection,\cite{brongersma_plasmon-induced_2015,brown_nonradiative_2016} radiative and resonance energy transfers.\cite{ma16}
However, while the relevance of topology was well established in electronic structure theory, \cite{bansil16} a little is known about purely topological shape effects in nanoplasmonics.
The notion of shape, standing in between geometry and topology, is used therein mostly in connection with the (differential) geometry of surfaces, e.g. SP in curved media, scattering and radiation at bends and interfaces.\cite{pfeiffer74,stockman04,veronis05,shitrit11}
Topology optimization of nanostructures\cite{jensen11} can be regarded as well as an unrestricted shape optimization of a geometric functional.\\
A primary shape effect in LSP resonance may be detected by the so-called figure of merit, the ratio between enhanced local and incident fields, whose formal dependence on the real and imaginary parts of the complex dielectric function is changing with the particle shape.\cite{west10}
The linear optical response then may be derived from Drude's model or some semiclassical extension of it,\cite{bassani75,mezzasalma18} aiming to include the relevant electron relaxation and transition processes implied by the metal band structure.
Such features, clearly, are still material- and chemistry-mediated, with the problem that electronic Schr\"{o}dinger's equations for particles of arbitrary shape may demand explicit solutions for very large quantum numbers.\\
On the other hand, since the time of De Witt and da Costa,\cite{dewitt57,dacosta81} surface geometry is known to affect the motion both of classical waves and quantum objects.
A number of studies\cite{ikegami92,cantele00,marchi05} were foreseeing the existence of curvature-driven topology eigenstates, inducing quantum interference phenomena and influencing surface charge transmission in/out a nanostructure.
As an example, when optical waves squeeze onto a curved thin dielectric layer (a film waveguide), a frictional energy of geometric nature was deduced in the wave-optics domain to act either as a potential well or barrier, according to the sign of surface concavity.\cite{batz08,valle10}
On this basis, it seems to be plausible that damping sources of topological origin may partake in the energy transfer elicited by nanoparticles (plasmonic hot carriers, photocatalytic processes, etc.).\\
In this work, to focus on the relation between topological shape effects, geometry and optical materials properties, a (reversible) sphere-to-cube plasmonic transition is thoroughly described for Au nanoparticles.
Spheres and semicubes (e.g. cubes with more or less rounded caps) were prepared by processing and quantifying the particle curvature at corners (vertices) and edges.
LSP spectra of all nanoparticle samples then were detected.
Theoretically, an effective yet abstract approach, to separate pure shape from materials effects, will be based on the framework of secular equations for geometric eigenmodes, quite recently brought to the attention of nanoplasmonics community by Garc\'ia de Abajo.\cite{deabaco97,deabaco02}
They define the complete basis set of naturally self-sustained multipoles with no external sources, in which the inhomogeneous term of the full electrostatic equation can be represented.
An analogy then was put forward (to the best of our knowledge for the first time) between the effect of shape and crystal (or ligand) field theory (CFT), ultimately resting on topology and group theory.
The theoretical section thus devises a formalism which then is applied to semicubic particles, here forming the majority of experimental samples.
To employ CFT, either in a retarded or non-retarded (quasistatic) picture, we were required to figure out in detail which effects may influence the LSP response.
Phenomena that are normally neglected or deemed as spurious (e.g. electrochemical) came here under close scrutiny, identifying numerically an excess charge correction which turns out to be necessary for a self-consistent description of experimental (retarded) data.

\begin{figure}[h]
	\centering
	\includegraphics[width=7.5 cm]{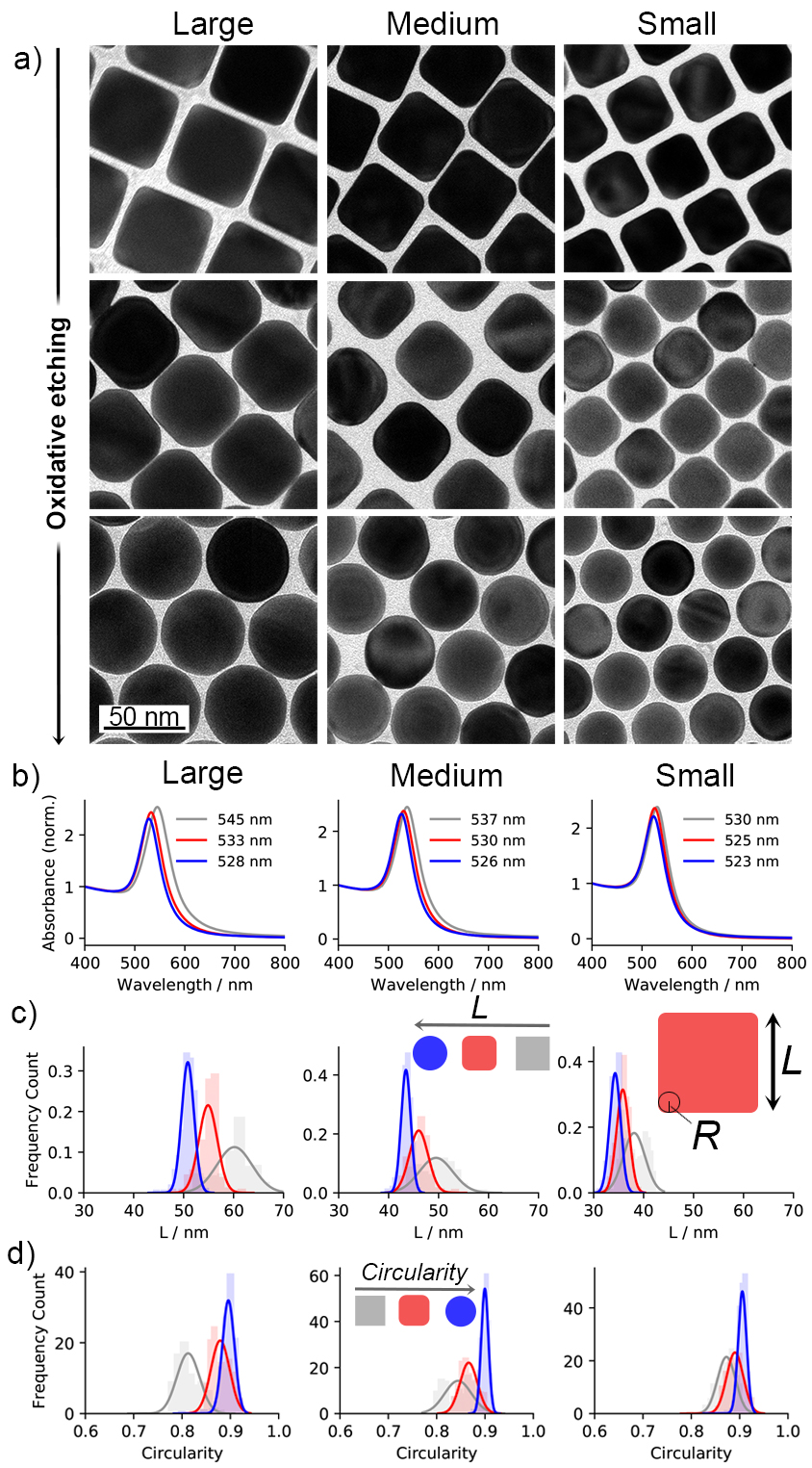}
	\caption{Shape transition from cubes to spheres for large, medium and small Au nanoparticles. a) TEM images of initial Au nanocubes (top row), semicubes obtained after mild oxidation (middle row), and spheres after moderate oxidation (bottom row). b) UV-Vis-NIR spectra of all nanoparticles showing gradual LSP blueshift upon oxidative etching. c) All length distributions versus oxidative etching. d) Circularity change for large, medium and small nanoparticles showing that circularity values increase upon the cube-to-sphere transition.}
	\label{fig:exp}
\end{figure}

\noindent {\rm \bf Oxidative etching-driven shape variation}. Because this work deals with a novel theoretical framework for plasmonic shape effects, it is of the utmost importance resorting to a synthesis scheme which is able of tuning shape variations with high accuracy.
We commenced our study with the synthesis of cube-like Au nanoparticles ($40$, $50$, and $60$ nm sized) that were subjected towards oxidative etching in the presence of Au salt, where the extent of etching is adjusted by changing the amount of Au precursor.\cite{rodriguez-fernandez_spatially-directed_2005}
This model drives oxidation to preferentially occur at metal surface sites with larger curvature, producing shape variations independently of the particle size.
In fact, since surface reactivity depends on curvature, the oxidation of cubes is favoured at the edges, causing their gradual rounding to eventually produce nearly spherical nanoparticles. 
To our aim, we selected the initial nanocubes, the partially oxidized cubes on edges, and fully rounded nanoparticles, which are referred hereinafter as cubes, semicubes and spheres, respectively (Figure \ref{fig:exp}a).
An ad-hoc analysis of electron microscopy images, counting at least $200$ particles each (see the last paragraph in the first Supporting Information - SI 1A), confirmed that the mean particle length $\langle L \rangle$ decreases upon the oxidation process (Figure \ref{fig:exp}c).
Interestingly, the extent of size reduction upon etching is more pronounced for larger particles, which is due to the well-defined cube-like morphology (higher shape anisotropy) of larger cubes, as compared to the smaller ones, displaying more rounded edges.
Worth mentioning is the fact that, upon oxidation, the size distribution get narrower, driving the particle monodispersity to increase with decreasing shape anisotropy.
In summary, the circularity value (${\rm C}$) was determined in each sample (Figure \ref{fig:exp}d, SI 2, Table (\ref{tbl:tab2})) and a further statistical analysis was performed to get the best predictions for the radius of spherical caps $\langle R \rangle$ and the shape descriptor used throughout this study, defined as ${\rm S} = 1 - \tfrac{2 \langle R \rangle}{\langle L \rangle}$.\\
Optical characterization revealed the oxidation process blueshifts the LSP band, in accord with previous studies (see e.g. ref.\cite{gen16}). 
Overall, it is more pronounced for larger than for smaller particles: from $545$ to $528$ nm, $537$ to $526$ nm, and $530$ to $523$ nm for large, medium and small particles, respectively (Figure \ref{fig:exp}b).
The pronounced blueshift for the largest particles well relates with their high shape anisotropy, and in turn with their variation in the length distribution (Figure \ref{fig:exp}c). 

\noindent
The summary of experimental values for the LSP peak wavelength ($\lambdabar$) is in Table (\ref{tbl:tab1}) of the next section, along with the relevant numerical predictions.
Chemical synthesis methods are in SI (1A).

\begin{figure}[h]
\centering
\includegraphics[scale=0.05]{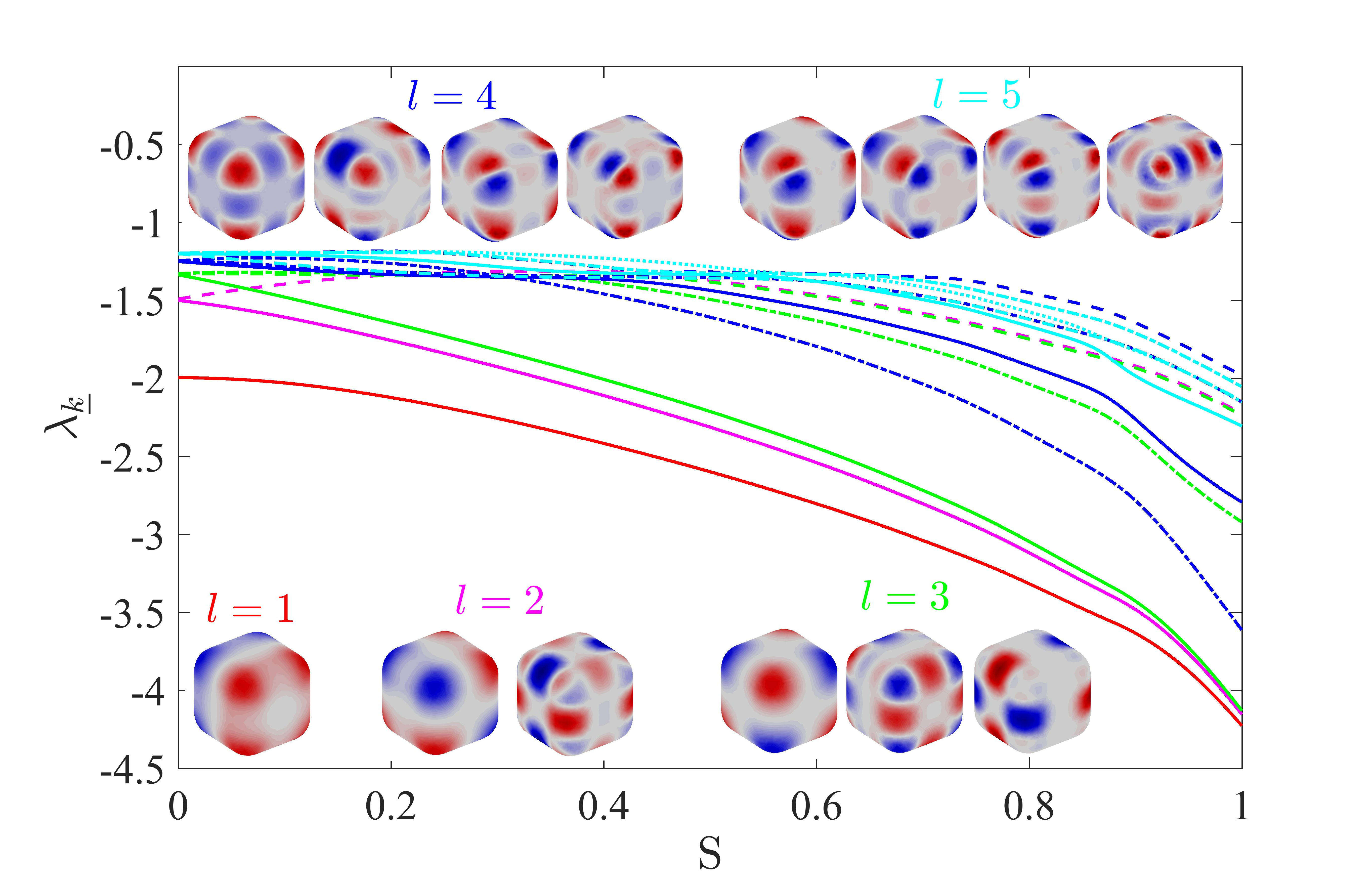}
	\caption{Eigenvalues ($1 \le \ell \le 5$) as a function of shape descriptor (${\rm S} = 0, 1$ for perfect spheres and cubes, respectively).
	Bands at equal $\ell$ were plotted with the same color.
	A space representation of all modes is also reported for the semicube with ${\rm S} = \tfrac{1}{2}$, their degeneracy conforming to equations (\ref{eqn:8b}-\ref{eqn:9b}).}
	\label{fig:eigenmodes}
\end{figure}

\noindent {\rm \bf Geometric normal modes}. In a quasistatic picture, neglecting retardation but preserving a frequency-dependent dielectric function ($\epsilon$), surface charges follow from the applied and induced external fields.\cite{deabaco02,Mayergoyz2005}
With no external sources, they are governed by the secular equation (\ref{eqn:1}) of the next Theoretical section.
Normal modes are characterized by a family of eigenvalues ($\lambda_{\underline{k}}$) and eigenfunctions ($\sigma_{\underline{k}}$) that only depend on the geometric shape and can be represented by a multipole expansion with Fourier's coefficients $\underline{k} = \{ \ell, m \}$.
Figure (\ref{fig:eigenmodes}) shows their evolution ($\ell \le 5$) when a sphere is transformed into a cube by varying the shape descriptor ${\rm S}$.
Between the extremes, (sphere) $0 < {\rm S} < 1$ (cube), lies a continuous class of semicubes with cylindrical contours, uniquely defined by the value of ${\rm S}$.
Numerical simulations were performed by the MNPBEM code,\cite{hohenester2012mnpbem} relying on the boundary element method (BEM) (see SI 1B)).\cite{de1998relativistic,deabaco02}
It returned a discrete set of $\sum (2 \ell + 1) = 35$ normal modes, corresponding to the most stable states.
For a sphere, they are given by spherical multipoles, each with degeneracy $2 \ell + 1$ ($m = - \ell$ ... $\ell$, see next equation \ref{eqn:6}).
Their individual evolution for ${\rm S} \neq 0$ shows a degeneracy breaking and defines in the end a sort of band structure.
However, tracking the mode evolution at a fixed $\ell$ is not supplied in MNPBEN by a direct toolbox output.
As eigenvalues are ordered by their value, and not by the multipole degree, it was necessary to introduce some analytical criterion to discriminate among them.
We found that imposing the numerical continuity to eigenvalues and their first derivative with respect to ${\rm S}$ was very efficient both to sort out different bands and to follow individual modes.
Results were clearly reversible, i.e. fully reproducible upon reverting the transition, i.e. ${\rm S} = 0 \rightarrow 1$ or ${\rm S} = 1 \rightarrow 0$.
In view to quantitatively describe, in the next section, degeneracy breaking and evolution of eigenvalues, it may be useful to comment on the first normal modes.
Dipole ($\ell = 1$) is the only one always retaining the same degeneracy.
The quadrupole of the sphere ($\ell = 2$) splits into a three-fold corner mode and a two-fold edge mode.
Octupole ($\ell = 3$) is sent into three bands, a non-degenerate corner mode, a three-fold edge mode, and a three-fold corner-edge mode.
Sphere eigenvalues monotonically increase with increasing multipole degree, whereas the three most stable cube modes are corner-like,\cite{nicoletti2013three} displaying dipolar, quadrupolar and octupolar nature.
Some visual representations may be found in Figure (\ref{fig:eigenmodes}) and in SI (3).
SI (1B) reports further details of the employed numerical methods.

\noindent {\rm \bf Predicting the optical spectra}. We used the MNPBEM toolbox under quasistatic and fully retarded approximation schemes.
The extinction cross sections followed from adopting the particle size and shape descriptor that came from the statistical analysis in SI (2A-C), correcting the initial circularity values from geometric error sources.
Optical response, which is isotropic as a consequence of geometric symmetry, turns out to be mostly dominated by dipole-like corner modes.
This is not, clearly, a rule of thumb as for example Ag nanoparticles show larger higher-order contributions to LSP spectra (SI 4).\\
Simulated data appeared to be systematically red-shifted by a number of nanometers in comparison with experiments.
In general, the larger and less spherical the particles were, the larger the shift magnitude was.
As proved in SI (5D-E), both the electrolyte-induced reduction of the solution dielectric function and polydispersity effects can be ruled out as possible causes for this anomalous red-shift.
A suitable explanation can be given instead by some excess charge at the Au/solution interface, implying a larger effective plasma frequency for Au particles that blue-shifts the plasmon resonance.
Surface charge mechanisms result from a complex interplay of intermolecular forces and can be well quantified by thorough thermodynamic models.\cite{mezzasalma96}
SI (5F) shows that the shift of the ratio between effective and bulk plasma frequencies may be either described by a linear dependence in the particle specific surface or in the shape descriptor.
By this linear law, and a standard procedure endowed with electronic interband transitions,\cite{bohr77} the Au dielectric function was recalculated with corrected plasma frequencies.
The final comparison among experiments and computations is depicted in Table (\ref{tbl:tab1}), proving a satisfactory agreement.
A synopsis of all experimental and numerical results is reported instead in Table (\ref{tbl:syn}) of SI (5G), where it may be seen that quasistatic predictions exceed the experimental values and turn out to be very close to corrected-retarded data (except for ${\rm S} = 0.81$).
While this is confirming the need to correct retardation here, a quantitative explanation of this occurrence is not yet ready and is left for future work.
Qualitatively, an available excess charge may tend to further balance the field across the particle.

\begin{table*}[!h]
\small
  \caption{Summary of LSP resonance predictions. Theoretical (${\rm th}$) and experimental (${\rm ex}$) wavelengths $\lambdabar$ are in nm.}
  \label{tbl:tab1}
  \begin{tabular*}{\textwidth}{@{\extracolsep{\fill}}lll|lll|lll}
    \hline
		 ${\rm S}$ & $\lambdabar$ \; (${\rm ex}$) & $\lambdabar$ \; (${\rm th}$) &
		 ${\rm S}$ & $\lambdabar$ \; (${\rm ex}$) & $\lambdabar$ \; (${\rm th}$) &
		 ${\rm S}$ & $\lambdabar$ \; (${\rm ex}$) & $\lambdabar$ \; (${\rm th}$) \\
		 \hline
		 0.51 & 530 & 531 & 0.67 & 537 & 540 & 0.81 & 545 & 557\\
		 0.41 & 525 & 528 & 0.58 & 530 & 534 & 0.55 & 533 & 534\\
		 0.26 & 523 & 524 & 0.38 & 526 & 524 & 0.44 & 528 & 527\\
    \hline
  \end{tabular*}
\end{table*}

\begin{figure*}[h]
\centering 
\includegraphics[width=\textwidth,height=15.4cm]{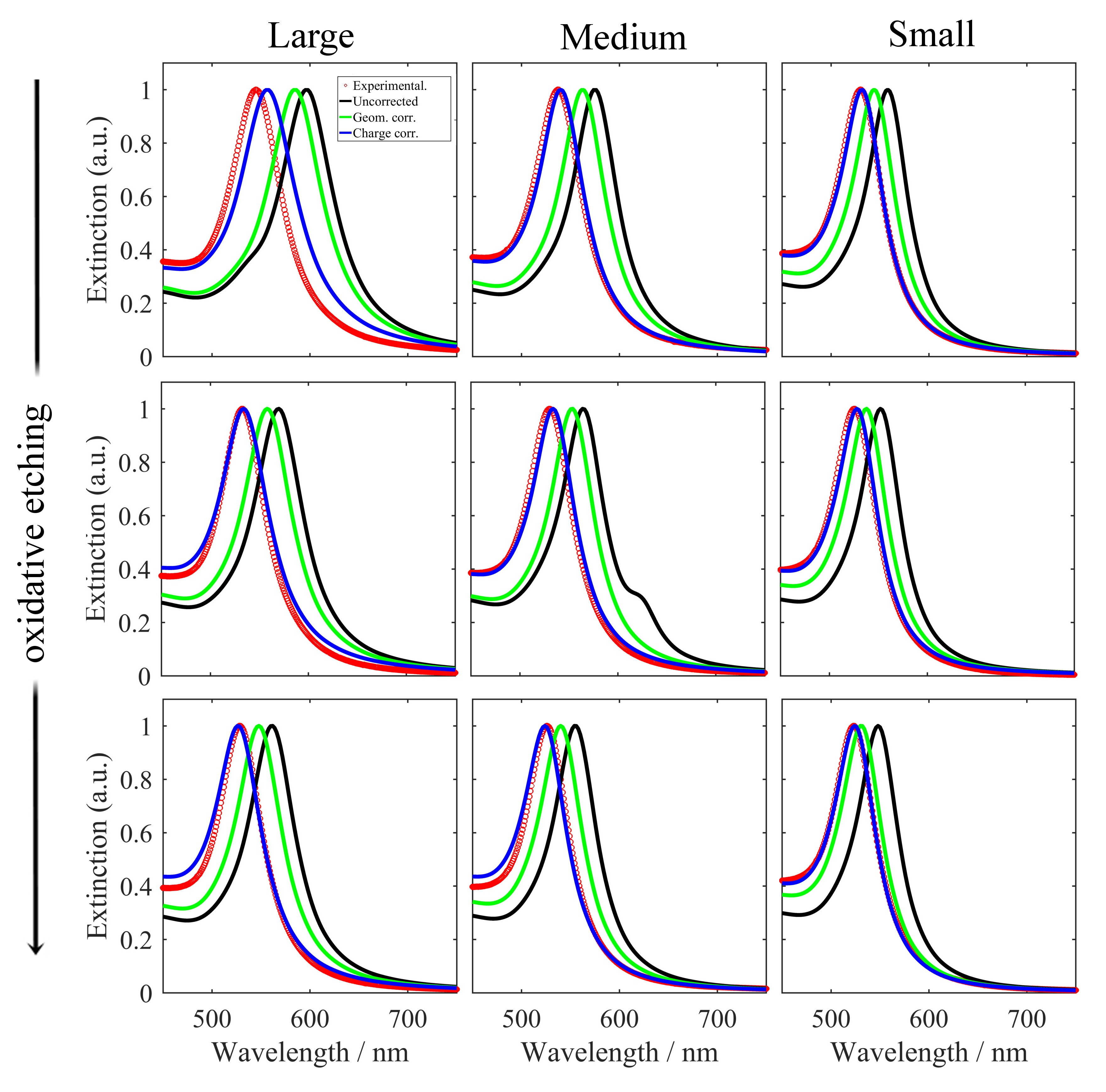}
	\caption{Experimental and simulated optical extinction spectra for all samples, corresponding to the TEM images in Figure (\ref{fig:exp}).
	Simulations with uncorrected (black), geometrically corrected (green) and charge corrected (blue) are superimposed to the experimental data (red) to show the proper trend of the approximations made and the accuracy of the final prediction.}
	\label{fig:fig3}
\end{figure*}

\noindent Figure (\ref{fig:fig3}) illustrates the uncorrected and corrected spectra (without and with excess charge) versus measurements.
The excellent agreement for the simulated widths confirms as well the accuracy of the low-loss imaginary part of the adopted dielectric function (Au single crystals).\cite{olmon2012optical} 
If other widely used optical constants were assumed,\cite{johnson1972optical} the predicted width would be larger than the experimental one.
To prove this conclusion, SI (5F) compares the results from single crystal and Johnson \& Christy's optical data for Au. 

\noindent {\rm \bf Secular equation}. The study of electrostatic normal modes, for a metal particle of arbitrary shape ($p$) in a dielectric medium ($d$), is basically a boundary condition problem, where eigenvalues are expressible by the ratio of the frequency-dependent dielectric constants, $\lambda \equiv \epsilon_p (\omega) / \epsilon_d (\omega)$.
The condition $\lambda_{\underline{k}} < 0$ translates the self-sustainability of natural oscillations, the energy maintaining the field externally being balanced by an equal amount of internal work, or vice versa.\cite{langbein75}
This limiting constraint identifies topology spectra that are only a signature of shape as, in the long-wavelength limit, retardation is negligible and transverse modes don't contribute to absorption.\\
Formally, surface modes take place when the polarization field inside the particle obeys the joint longitudinal and transverse condition, $\nabla \cdot {\bf P} = \nabla \wedge {\bf P} = 0$, while a polarization charge $\nabla \cdot {\bf P} \neq 0$ develops at the surface.
In the language of potential theory, equivalent double-layer or single layer representations may be afforded for the surface charge density $\sigma$:\cite{ouyang89}
\be
\int \sigma ({\bf r}') \p_{\bf n} \f{1}{|{\bf r} - {\bf r}'|} d s' \; = \; \Lambda_q \sigma ({\bf r})
\label{eqn:1}
\ee
or for the internal electric potential $V_i$:\cite{vangelder72}
\be
\int \f{\p_{\bf n} V_i ({\bf r}')}{|{\bf r} - {\bf r}'|} d s' \; = \; \Lambda_v V_i ({\bf r})
\label{eqn:2}
\ee
$\p_{\bf n}$ being the normal derivative at the surface.
These integrals include the static limit of Green's function $|{\bf r} - {\bf r}'|^{-1}$ with observation point ${\bf r}$, and the (real-valued) eigenvalues $\Lambda_q = 2 \pi (1 + \lambda)/(1 - \lambda)$ and $\Lambda_v = 4 \pi/(1 - \lambda)$ for the two problems, which then can be related to the secular equation in frequency domain provided an optical dispersion law is stated for the dielectric function.

\noindent {\rm \bf Sphere eigenvalues from vanishing electrostatic energy}. The last observations hold when Mie's theory is applied to a sphere, getting for a given plasma frequency ($\omega_{\rm p}$) and angular momentum ($\ell$) of a plasmon mode, $\omega^2_\ell / \omega^2_{\rm p} = \ell/ (2\ell + 1)$.\cite{hohen05}
Drude's model then allows to recover:
\be
\lambda_{\underline{k} s} \; = \; - 1 - 1/\ell \;\;\;\;\;\;\;\; ({\rm sphere}) 
\label{eqn:6}
\ee
We now prove by another way that equation ({\ref{eqn:6}) gives the eigenvalues of spherical harmonics $Y_{\ell m} (\theta, \varphi)$ ($\theta$, $\varphi$ $=$ polar, azimuth angles) at a vanishing electrostatic energy ($E = 0$) in the flux equation for the fields internal ($i$) and external ($e$) to the particle:\cite{langbein75}
\be
8 \pi E \; = \; \epsilon_p \int V_i ({\bf r}) \nabla V_i \cdot d {\bf s} \; - \; \epsilon_d \int V_e ({\bf r}) \nabla V_e \cdot d {\bf s} 
\ee
i.e.:
\be
\lambda \; = \; \f{\int V_e \p_{\bf n} V_e d s}{\int V_i \p_{\bf n} V_i d s}
\label{eqn:3}
\ee
To test equation (\ref{eqn:3}), which will be fruitful to deal with semicubes, a comparison with equation (\ref{eqn:1}) is in order.
In an extended formulation, one should find two Green functions, defined at both surface sides for each mode $\underline{k}$, giving a simultaneous solution for the two problems, i.e. let ${\rm G}_h ({\bf r}, {\bf r}') = 1/|{\bf r} - {\bf r}'|_h$ ($h = i$, $e$), fulfilling:
\be
\p_{\rm n} \f{1}{|{\bf r} - {\bf r}'|} \; = \; \tfrac{1}{2} \left( \p_{\rm n} {\rm G}_i + \p_{\rm n} {\rm G}_e \right)
\label{eqn:4}
\ee
and, for any constant $R \in \Re^+$:
\be
\lambda_{\underline{k} s} \; = \; \lim_{r,r' \rightarrow \; R} \f{\iint {\rm G}_e \p_{\bf n} {\rm G}_e d s d s'}{\iint {\rm G}_i \p_{\bf n} {\rm G}_i d s d s'}
\label{eqn:7}
\ee
We start from Green's function expansion:
\be
r_{_>} {\rm G}_h ({\bf r},{\bf r}') \; = \; \sum^{\infty}_{\ell = 0} \sum^{\ell}_{m = - \ell}
\f{4 \pi}{2 \ell + 1} \left( \f{r_{_<}}{r_{_>}} \right)^\ell Y^*_{\ell m} (\theta', \varphi') Y_{\ell m} (\theta, \varphi)
\ee
with $r_{_<} = {\rm min} \{r , r'\}$, $r_{_>} = {\rm max} \{r , r'\}$, and consider a sphere of radius $r$, i.e. ${\rm G}_i (r_{_<}, r_{_>}) = {\rm G}_i (r', r)$ and ${\rm G}_e (r_{_<}, r_{_>}) = {\rm G}_e (r, r')$.
If we define the radial terms per mode $\ell$ in equation (\ref{eqn:4}) as ${\rm g}_h (r_{_<}, r_{_>}) \equiv \p_{\rm n} {\rm G}_h (r, r')$:
\be
{\rm g}_i (r', r; \ell) = \f{\ell'}{r r'} \left( \f{r'}{r} \right)^{\ell'} \delta_{\ell \ell'} , \;\;\;
{\rm g}_e (r, r'; \ell) = - \; \f{1 + \ell'}{r'^2} \left( \f{r}{r'} \right)^{\ell'} \delta_{\ell \ell'}
\label{eqn:delta}
\ee
then for any spherical element of radius $R$, the basic term in equation (\ref{eqn:7}) may be written in a short-handed notation as:
\be
\int {\rm G}_h \p_{\bf n} {\rm G}_h d s = \f{(4 \pi)^2}{R} \sum_{\ell' \ell'' m' m''} \f{\gamma_{h \ell'} \delta_{\ell \ell'}}{(2 \ell' + 1)(2 \ell'' + 1)} \int \Upsilon_{\ell' \ell'' m' m''} d \Omega
\label{eqn:5}
\ee
with:
\be
\Upsilon_{\ell' \ell'' m' m''} = \; Y^*_{\ell' m'} (\Omega') Y_{\ell' m'} (\Omega) Y^*_{\ell'' m''} (\Omega') Y_{\ell'' m''} (\Omega)
\ee
and:
\be
\gamma_{h \ell'} \; \delta_{\ell \ell'} \; = \; R^{2} \lim_{r, r' \rightarrow R} {\rm g}_h \; (\ell)  
\ee
$d \Omega$ being the solid angle element subtended by $d s$ at the observation point.
From the properties of spherical harmonics, we get:
\be
\lim_{r,r' \rightarrow \; R} 
\iint {\rm G}_h \p_{\bf n} {\rm G}_k d s d s' \; = \; \f{(4 \pi)^2 R}{2 \ell + 1} \; 
\gamma_{h \ell}
\label{eqn:mode}
\ee
and thus the eigenvalue equation (\ref{eqn:6}), as $\gamma_{i \ell} = \ell$ and $\gamma_{e \ell} = - \; 1 - \ell$.\\
To complete the proof, one should ascertain equation (\ref{eqn:4}) is consistent with the secular equation.
In this case, the average normal derivative is independent of $\ell$:
\be
\tfrac{1}{2} \lim_{r,r' \rightarrow R} ({\rm g}_i + {\rm g}_e) \; = \; - \; \tfrac{1}{2} R^{- 2} 
\ee
and equation (\ref{eqn:1}) can be written as:
\be
- \; 2 \pi \sum^{\infty}_{\ell' = 0} \sum^{\ell'}_{m' = - \ell'} \f{Y_{\ell' m'} (\Omega)}{2 \ell' + 1} \int \sigma_{\underline{k}} ({\bf r}') Y^*_{\ell' m'} (\Omega') d \Omega' \; = \; \Lambda_{\underline{k} q} \; \sigma_{\underline{k}} ({\bf r})
\label{eqn:8}
\ee
It is clear that any eigenfunction $\sigma_{\underline{k}} ({\bf r}) = \sigma_{_0} Y_{\ell m} (\Omega)$ will solve equation (\ref{eqn:8}) whenever $\sigma_{_0}$ is a constant charge density and, from the spherical harmonics normalization:
\be
- \; 2 \pi \sum^{\infty}_{\ell' = 0} \sum^{\ell'}_{m' = - \ell'} \delta_{\ell \ell'} \delta_{m m'} \f{Y_{\ell' m'} (\Omega)}{2 \ell' + 1} \; = \; \Lambda_{\underline{k} q} Y_{\ell m} (\Omega)
\ee
eigenvalues depend on angular momenta like $\Lambda_{\underline{k} q} = - 2 \pi / (2 \ell + 1)$, bringing back to equation (\ref{eqn:6}).
Observe, for $\lambda < 0$ to be attained, the signs of normal gradients should be opposite, no matter which specific convention may be adopted.
Accordingly, forthcoming applications of equation (\ref{eqn:3}) will stick for simplicity to equations (\ref{eqn:delta}) ($\p_{\rm n} {\rm G}_i > 0$ and $\p_{\rm n} {\rm G}_e < 0$).

\noindent {\rm \bf Plasmon resonance in semicubes from non-retarded crystal field splitting}. Deforming a sphere into a highly spherical semicube (${\rm S} \rightarrow 0^+$) reduces the particle symmetry, leading eigenvalues to a degeneracy removal.\cite{wol77}
The phenomenology subtended by Figure (\ref{fig:eigenmodes}) will be explained by CFT, pioneered long ago by Bethe\cite{bet29} to model coordination complexes.
Evolution of edges, faces and corners will be described by a minimal model of effective surface charges, mimicking the topology transition by a suitable charge redistribution.\\
To fix the ideas, let a ${\rm S} = 0 \rightarrow 0^+$.
At a given $\underline{k} = (\ell, m)$, each eigenvalue $\lambda_{\underline{k}} = \lambda_{\underline{k}} ({\rm S})$ can be mapped into a monotonically increasing energy function by some optical dispersion model, i.e.:
\be
\varepsilon^{-2}_{\underline{k}} = \; 1 - \lambda_{\underline{k}} \;\;\;\;\;\;\;\; ({\rm Drude's})
\label{eqn:8c}
\ee
Energy will vary with respect to the barycenter values $\varepsilon_{\underline{k} s} \equiv \varepsilon (\lambda_{\underline{k} s})$ of the set of spherical multipoles, i.e. $\varepsilon_{\underline{k} s} \rightarrow \varepsilon_{\underline{k} +}$.
We can model the topology change by introducing some excess charge at the face centers of the cube, with ${\rm S} = 1$ taken on as a reference state.
This makes our problem equivalent to the crystal field splitting of $1$-electron orbitals in the cubic-symmetry potential of an octahedral disposition with point group $O_h$.\cite{figgis00} Trends of eigenvalues upon ${\rm S} = 0 \rightarrow 0^+$ confirms such a spectroscopic analogy.
First, the $x$, $y$, $z$ symmetries of $P$ states ($\ell = 1$) disallow a degeneracy removal:
\be
\varepsilon_{_+} \{ x_k \} = \varepsilon_{_+} \{ x_h \} \;\;\;\;\;\;\;\;\;\; (P)
\label{eqn:8b}
\ee
while enquiring the cases $\ell = 2$, $3$, $4$ allows to identify the following phenomenological series:
\be
\begin{aligned}
\varepsilon_{_+} & \{ x_k x_h \} < \varepsilon_{_+} \{ x^2 - y^2, z^2 \} \;\;\;\;\; (D)\\
\varepsilon_{_+} & \{ x y z \} < \varepsilon_{_+} \{ x_k (x^2_h - x^2_q) \} < \varepsilon_{_+} \{ x_k (x^2_k - r^2) \} \;\;\;\;\; (F)\\
\varepsilon_{_+} & \{ (z^2-r^2)(x^2-y^2), (x^2-r^2)(y^2-z^2)-(y^2-r^2)(z^2-x^2) \} \\
< & \; \varepsilon_{_+} \{ x_k x_h (x^2_k - x^2_h) \} < \varepsilon_{_+} \{ x^4 + y^4 + z^4 - r^4 \} \;\;\;\;\; (G)
\end{aligned}
\label{eqn:9}
\ee
subindices $\underline{k}$ and coefficients of orbital polynomials\cite{king97} being omitted for simplicity, with $x_k, x_h, x_q \in \{x, y, z \}$.\\
All of these relationships agree with the predictions for octahedral field splitting,\cite{scl69,per81} i.e. in Mullikan's notation:
\be
\begin{aligned}
\varepsilon \; ( T_{2 g} ) & < \varepsilon \; ( E_{g} ) & (D)\\
\varepsilon \; ( A_{2 g} ) & < \varepsilon \; ( T_{2 g} ) < \varepsilon \; (T_{1 g}) & (F)\\
\varepsilon \; ( T_{2 g} ) & < \varepsilon \; ( E_{g} ) < \varepsilon \; ( T_{1 g} ) < \varepsilon \; ( A_{1 g} ) & (G)
\end{aligned}
\label{eqn:9b}
\ee
where, in parentheses, are the states representative of irreducible representations of $O_h$, with degeneracy values that exactly correspond to equations (\ref{eqn:9}), i.e. ${\rm deg} (A_g) = 1$, ${\rm deg} (E_g) = 2$ and ${\rm deg} (T_g) = 3$.\cite{scl69}
The eleven states of $H$ series split into $2,3,3,3$ levels (from lower to higher energies), which agree with the symmetry group properties of atomic orbitals with $\ell = 5$.\cite{king97}
A pictorial representation of the electromagnetic analog of crystal field splitting is sketched out in Figure (\ref{fig:TOC}) (see also Tanabe-Sugano diagrams).

\begin{figure}[h]
	\centering
	\includegraphics[scale=0.32]{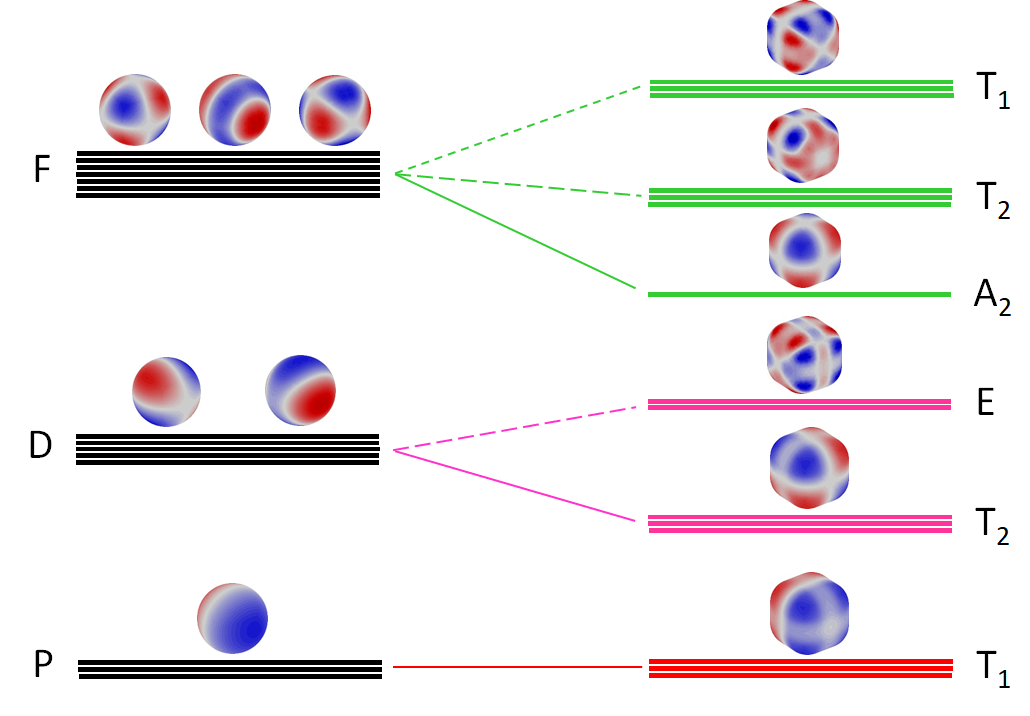}
	\caption{Scheme of the plasmon splitting mechanism, in a potential of octahedral symmetry ($O_h$), implied by CFT (image not in energy scale, colors as in Fig. \ref{fig:eigenmodes}).}
	\label{fig:TOC}
\end{figure}	

\noindent At larger ${\rm S}$ values, $\lambda_{{\underline{k}}}$ are not monotonic thereby reflecting a more general symmetry behavior.
This is not surprising, as the tetrahedral $T_d$ is another relevant cubic group in our context.
While degeneracy of $P$ states won't still be removed upon $T_d$, an extended law should explain the increasing trend of $\lambda_{{\underline{k}}}$ with increasing ${\rm S}$ ($\ell > 1$).
As it is usual in CFT, equation (\ref{eqn:3}) may be generalized by projecting the potential onto the implied bra-ket eigenmodes.
$V_i$, thus, is ascribed to the spherically symmetric monopole produced at a radius $L/2$ by an innermost pointwise charge $q_i$, e.g. in Gaussian units:
\be
\langle \; 0 0 \; | \; V_i \p_{\bf n} V_i \; | \; 0 0 \; \rangle \equiv \tfrac{1}{2} \int Y^*_{_{0 0}} Y_{_{0 0}} \p_{\bf n} (V^*_i V_i) d s \; = \; 2 \f{q_i^2}{L} 
\label{eqn:13b}
\ee
where the complex conjugate of $V_i$ guarantees a real-valued functional and the convention on the normal derivative was taken in conformity to equations (\ref{eqn:delta}).
To get the numerator in equation (\ref{eqn:3}), we remind the most general form the real part of octahedral, tetrahedral and cubic potentials assume,\cite{figgis00} and write (SI 6H):
\be
V_{\ell} ({\bf r}) \; = \; \f{q_{e}}{r} \sum^{2 \ell}_{\ell' \; \in \; 2 {\rm N} \setminus 2} \o{r^{\ell'}} \f{{\rm L}_{\ell \ell'}}{r^{\ell'}} \;\;\;\; (\ell \ge 1)
\label{eqn:10b}
\ee
Here, $q_e$ is the effective charge generating $V_e$, the spatial average in the sum is performed over a radial wave function ($\o{r^{0}} = 1$) and ${\rm L}_{\ell \ell'}$ is a linear superposition of the form:
\be
{\rm L}_{\ell \ell'} = \; {\rm L}_{\ell \ell'} \{ Y_{\ell' 0}, Y_{\ell' \pm 4} \}
\ee
Then we consider $V_e = V_\ell$ into:
\be
\begin{aligned}
& \langle \; X_{\underline{k}} \; | \; V_e \p_{\bf n} V_e \; | \; X_{\underline{k}} \; \rangle = \tfrac{1}{2} \int q^2_e X^*_{\ell m} \p_{\bf n} \left[ ( \o{r^{0}} \f{{\rm L}^*_{\ell 0}}{r} + \o{r^{4}} \f{{\rm L}^*_{\ell 4}}{r^5} + \; \o{r^{6}} \f{{\rm L}^*_{\ell 6}}{r^7} \; ... \; ) \right. \\
& \left. ( \o{r^{0}} \f{{\rm L}_{\ell 0}}{r} + \o{r^{4}} \f{{\rm L}_{\ell 4}}{r^5} + + \; \o{r^{6}} \f{{\rm L}_{\ell 6}}{r^7} \; ... \; ) \right] X_{\underline{k}} \; d s
\end{aligned}
\label{eqn:10}
\ee
that is clearly truncated to $\ell' \le \ell$ and, from symmetry arguments, doesn't display any term for $\ell' = 2$ (${\rm L}_{\ell 2} \equiv 0$).\cite{scl69}
Cubic harmonics $X_{\underline{k}}$, i.e. linear combinations of real $Y_{\ell m}$ at fixed $\ell$, form the natural basis in irreducible representations of cubic groups.\\
To highlight the role of shape descriptor when an arbitrary surface is concerned, we resort to the mean value theorem for definite integrals (SI 6I).
The particle is divided into cubic ($c$), spherical ($s$) and semicubic ($b$) domains, each contributing (in units of $6 L^2$) to the overall area as $a_{c} = S^2$, $a_{s} = \tfrac{\pi}{6} (1 - S)^2$, $a_{b} = \tfrac{\pi}{2} S (1 - S)$.
Let $f$ be a continuous function onto a compact and rectifiable surface, the following notation is employed:
\be
\f{1}{6 L^2} \int f d s \; = \; \langle f \rangle_{_c} a_{c} + \langle f \rangle_{_s} a_{s} + \langle f \rangle_{_b} a_{b}
\label{eqn:11}
\ee
the effective charge ($q_e$) variation from sphere to cube obeying a minimal, linear and homogeneous, description:
\be
q_{\underline{k} b} ({\rm S}) \; = \; q_{\underline{k} s} \; + \; (q_{\underline{k} c} - q_{\underline{k} s}) {\rm S}
\label{eqn:12}
\ee
In comparison to the perfect spherical symmetry (equation \ref{eqn:7}), with $q_i = q_e$ (${\rm S} = 0$ in the last equation), a change of shape in CFT is carrying a charge displacement.
From equations (\ref{eqn:13b}), (\ref{eqn:10}-\ref{eqn:12}), equation (\ref{eqn:3}) now may be written as:
\be
\lambda_{\underline{k} b} = \; - \; (z_{\underline{k} s} + z_{\underline{k} b} {\rm S})^2 \left[ \sigma_{\ell m} (1 - {\rm S})^2 + \mu_{\ell m} {\rm S} (1 - {\rm S}) + \chi_{\ell m} {\rm S}^2 \right]
\label{eqn:13}
\ee
where, for any mode, $z_{\underline{k} s}$ and $z_{\underline{k} b}$ express $q_{\underline{k} s}$ and $q_{\underline{k} c} - q_{\underline{k} s}$ in unit of charge $q_i$, and $\sigma_{\ell m}$, $\mu_{\ell m}$, $\chi_{\ell m} > 0$ are, respectively, averages of the spherical, semicubic and cubic contributions to $\lambda_{\underline{k} b}$, all depending on the mean values $\o{\varrho^{\ell'}}$, with $\varrho \equiv r/L$:
\be
\sigma_{\ell m} = - \tfrac{\pi}{4} \langle \; X_{\underline{k}} \; | \; \p_{\bf n} \; [ ( \o{\varrho^{0}} \f{{\rm L}^*_{\ell 0}}{\varrho} + \o{\varrho^{4}} \f{{\rm L}^*_{\ell 4}}{\varrho^5} ... \; ) ( \o{\varrho^{0}} \f{{\rm L}_{\ell 0}}{\varrho} + \o{\varrho^{4}} \f{{\rm L}_{\ell 4}}{\varrho^5} \; ... \; )] \; | \; X_{\underline{k}} \; \rangle_s
\ee
\be
\mu_{\ell m} = - \tfrac{3 \pi}{4} \langle \; X_{\underline{k}} \; | \; \p_{\bf n} \; [ ( \o{\varrho^{0}} \f{{\rm L}^*_{\ell 0}}{\varrho} + \o{\varrho^{4}} \f{{\rm L}^*_{\ell 4}}{\varrho^5} ... \; ) ( \o{\varrho^{0}} \f{{\rm L}_{\ell 0}}{\varrho} + \o{\varrho^{4}} \f{{\rm L}_{\ell 4}}{\varrho^5} \; ... \; )] \; | \; X_{\underline{k}} \; \rangle_b
\ee
\be
\chi_{\ell m} = - \tfrac{3}{2} \langle \; X_{\underline{k}} \; | \; \p_{\bf n} \; [ ( \o{\varrho^{0}} \f{{\rm L}^*_{\ell 0}}{\varrho} + \o{\varrho^{4}} \f{{\rm L}^*_{\ell 4}}{\varrho^5} ... \; ) ( \o{\varrho^{0}} \f{{\rm L}_{\ell 0}}{\varrho} + \o{\varrho^{4}} \f{{\rm L}_{\ell 4}}{\varrho^5} \; ... \; )] \; | \; X_{\underline{k}} \; \rangle_c
\ee
In the spirit of equation (\ref{eqn:11}), subindexes $s$, $b$, $c$ mean averaging over corners (spherical), edges (cylindrical) and faces (Cartesian).
Because of the tough calculations involved (Clebsh-Gordan coefficients and integration over arbitrary surfaces), their numerical determination falls beyond the aims of this study, but equation (\ref{eqn:13}) specifies a minimal fourth-degree polynomial, well fitting the eigenvalues in the whole ${\rm S}$ domain with $\chi_{\ell m}$, $\mu_{\ell m}$, $\sigma_{\ell m}$ $\in (0.4-1.3)$, $z_{\underline{k} s} \in (1-2)$, $z_{\underline{k} b} \in (0-1)$ (see SI 6L).
Such coefficients, got by Levenberg \& Marquardt's algorithm, are correlated with $\ell$ (negatively in $\chi_{\ell m}$, $\mu_{\ell m}$, $z_{\underline{k} s}$; positively otherwise) but poorly correlate with state degeneracies.
In general, the effective charge progressively redistributes upon ${\rm S} = 0^+ \rightarrow 1^-$ into semicubic ($\mu_{\ell m}$) and cubic ($\chi_{\ell m}$) contributions.
Observe the charge defect $z_{\underline{k} b}/z_{\underline{k} s} = q_{\underline{k} c}/q_{\underline{k} s} - 1$ here tends to be negligible in dipoles ($\le 5 \cdot 10^{-3}$) and quadrupoles ($\le 5 \cdot 10^{-2}$), prompting the topology contribution of dominant modes to be only slightly dependent on geometry, and justifying further the charge-surface factorization in equation (\ref{eqn:13}).

\noindent {\rm \bf Retarded analog of crystal field splitting}.
In a quasistatic picture, equation (\ref{eqn:13}) affords a topological interpretation of LSP resonance, as Drude's model predicts $\lambdabar \propto \sqrt{1 - \lambda}$ (see equation \ref{eqn:8c}).
When surface effects are discarded, then we expect from the only topological charge term:
\be
\f{\sqrt{\Delta \lambdabar^2}}{\lambdabar_{_0}} \; \approx \; \o{z}_s + \o{z}_b {\rm S}  \;\;\;\;\;\;\;\;\;\; ({\rm non-retarded})
\label{eqn:14}
\ee
where, at some wavelength scale $\lambdabar_{_0}$, one has $\Delta \lambdabar^2 \equiv \lambdabar^2 - \lambdabar^2_{_0}$, with $\lambdabar$ being the resonance wavelength.
Equation (\ref{eqn:14}) has been applied to quasistatic LSP data (column qs in Table 4 of SI 5G).
Indications from the analysis of multipole coefficients at different $(\ell, m)$ were accounted for by a couple of independent best fits, one constrained to the numerical ranges of $\o{z}_k$ and the second to a negligible charge defect in dipole and quadrupole modes ($\o{z}_b \approx 0$).
The two interpolations are shown in Figure (\ref{fig:lasfig}), both in satisfactory agreement with the computed values.
\begin{figure}[h]
	\centering
	\includegraphics[scale=0.0415]{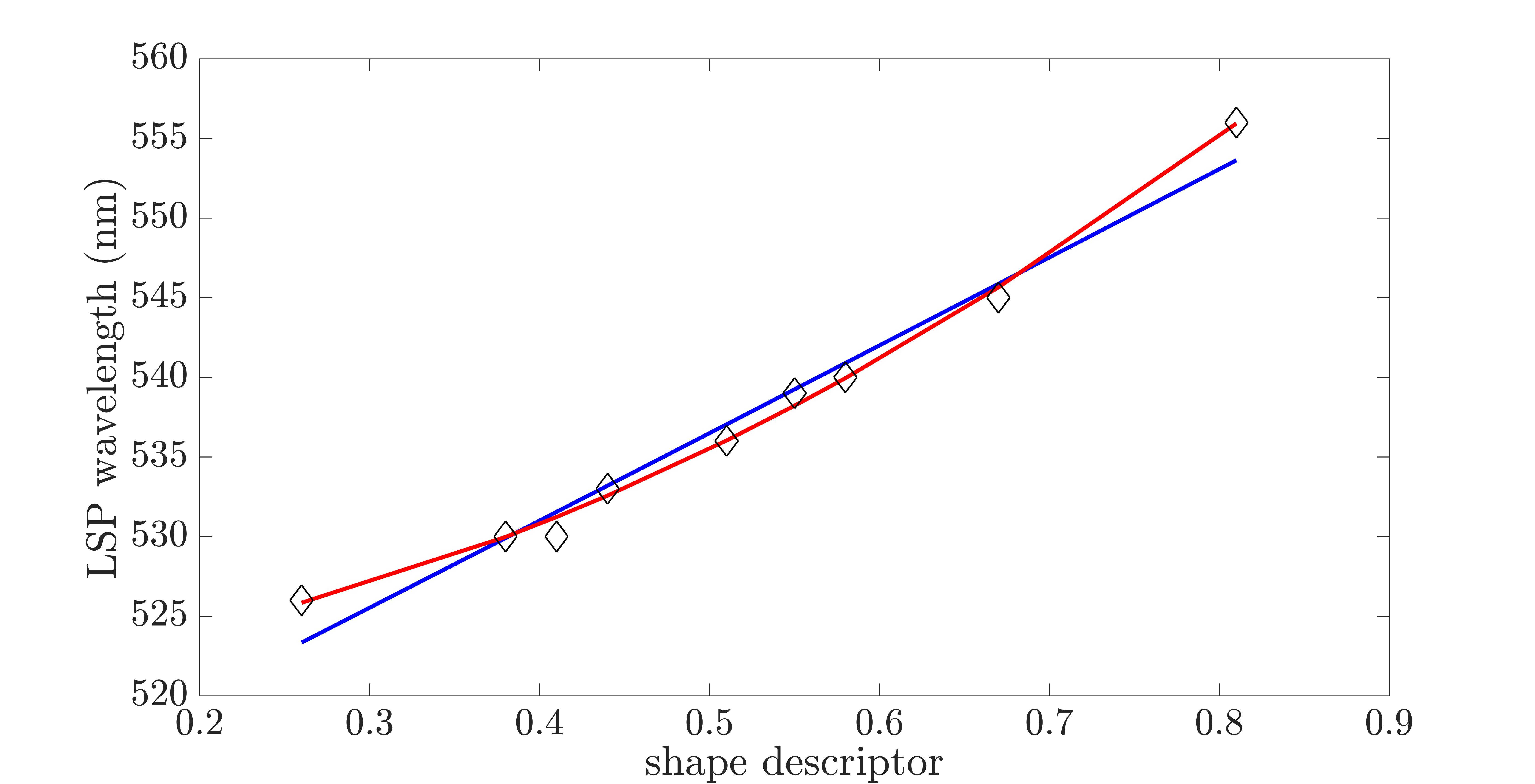}
	\caption{Quasistatic LSP wavelength $\lambdabar \; ({\rm qs})$ vs shape descriptor ${\rm S}$ (points). Blue and red lines (equation \ref{eqn:14}) respectively for $\o{z}_s = 1.65$, $\o{z}_b = 0.24$, $\lambdabar_{_0} = 236.7$ nm ($R^2 = 0.97$), and $\o{z}_s = 9.9 \cdot 10^{-3}$, $\o{z}_b = 0.44$, $\lambdabar_{_0} = 521.8$ nm ($R^2 = 0.99$).}
	\label{fig:lasfig}
\end{figure}
To include retardation in CFT by analytical means may be rather demanding, especially for non-spherical shapes.
However, with some judicious choice of retarded fields an evaluation of finite-size effects may be provided, even though remarkable corrections are not expected.
Spherical $\ell$-modes should get (meaningfully) retarded when $\ell < L / (2 \lambdabar)$,\cite{fuc68} which here is far from being satisfied.
To proceed, Coulomb's gauge for the vector potential, ${\bf \nabla} \cdot {\bf {\cal A}} = 0$, and the harmonic behavior in time $t$ of each field, e.g. ${\bf {\cal A}} ({\bf r}, t) = {\bf A} ({\bf r}) \; {\rm e}^{- {\rm i} \omega t}$, are adopted.
In this way the potential separates\cite{jack02} into an instantaneous (inst) and a retarded (ret) component, ${\bf A} = {\bf A}^{\rm inst} + {\bf A}^{\rm ret}$, being:
\be
{\bf A}^{\rm inst} = \; \f{\rm i}{k} \int \f{\rho ({\bf r}') \; \widehat{\bf R}}{|{\bf r} - {\bf r}'|^2} \; d^3 r' 
\ee
and, in turn, ${\bf A}^{\rm ret} = {\bf A}^{\rm ret}_{\bf J} + {\bf A}^{\rm ret}_{\rho}$, where the retarded potentials for a localized oscillating system of currents and charges are:
\be
{\bf A}^{\rm ret}_{\bf J} = \; \f{1}{c} \int \f{{\rm e}^{{\rm i} k |{\bf r} - {\bf r}'|}}{|{\bf r} - {\bf r}'|} \; {\bf J} ({\bf r}') d^3 r' 
\ee
\be
{\bf A}^{\rm ret}_{\rho} = \; \int \f{{\rm e}^{{\rm i} k |{\bf r} - {\bf r}'|}}{|{\bf r} - {\bf r}'|} \left ( \f{1}{{\rm i} k |{\bf r} - {\bf r}'|} - 1 \right) \rho ({\bf r}') \; \widehat{\bf R} \; d^3 r' \;,
\label{eqn:rho}
\ee
$\widehat{\bf R}$ is the unit vector in the direction pointed by ${\bf r} - {\bf r}'$ and $k = \omega / c$ is the wavenumber ($c =$ light speed).
${\bf J}$ and $\rho$ denote respectively the spatial parts of the harmonic current and charge densities, the latter producing the instantaneous crystal field potentials setting the eigenvalues in the integral equation (\ref{eqn:3}), i.e.:
\be
V_k ({\bf r}) \; = \; \int \f{\rho_{_k} ({\bf r}')}{|{\bf r} - {\bf r}'|} d^3 r' 
\ee
Correspondingly, the electric field ${\bf E}_k = - {\bf \nabla} V_k - \tfrac{1}{c} \tfrac{\p {\bf \cal A}}{\p t}$ writes:
\be
{\bf E}_k = \; {\rm i} k {\bf A}^{\rm ret}
\ee
the instantaneous gauge condition ${\rm i} k {\bf A}^{\rm inst} = {\bf \nabla} V_k$ being taken into account.
On these basis, equations (\ref{eqn:3}) and (\ref{eqn:7}) transform according to the condition $\p_{\bf n} V_k \rightarrow \p_{\bf n} V_k - {\rm i} k {\bf A} = - {\rm i} k {\bf A}^{\rm ret}$, which here should be evaluated in the intermediate (induction) domain.\\
Before entering the application to semicubes, an analysis of dipole corrections will be developed for spherical particles, proving that consistent results for $\ell = 1$ follow at third order in $k r$.
To this aim, the multipole expansion for ${\bf A}^{\rm ret}_{\bf J}$ is available at the transition from quasistatic to radiation zones, while the charge density potential can be expanded as (SI 6M):
\be
{\bf A}^{\rm ret}_{\rho e} \; = \; \f{\rm i}{k} {\bf \nabla} V_e \; + \; \delta {\bf A}^{\rm ret}_{\rho e}
\label{eqn:n}
\ee
with:
\be
\delta {\bf A}^{\rm ret}_{\rho e} \; = \; - \;
\sum_{n \; \ge \; 1} \f{n}{(n + 1)!} ({\rm i} k)^n {\bf I}_n
\ee
\be
{\bf I}_n = \int ({\bf r} - {\bf r}') |{\bf r} - {\bf r}'|^{n - 2} \rho ({\bf r}') \; d^3 r'
\ee
The purely geometrical equation (\ref{eqn:7}), which was able to recover the non-retarded spectrum equation (\ref{eqn:6}), is corrected now by the only first term of Lorenz' potential detected externally to the particle surface:\cite{jackb}
\be
{\bf A}^{\rm ret}_{\bf J e} = \; - \; \f{{\rm i} k}{r} {\rm e}^{{\rm i} k r} {\bf p}
\ee
where ${\bf p}$ denotes the electric dipole moment.
Let the shorthand notation $a_{\bf n} \equiv {\bf n} \cdot {\bf a}$ and keeping for the moment the first term ($n \equiv 0$) in equation (\ref{eqn:n}), it turns out:
\be
- \; {\rm i} k A^{\rm ret}_{\bf n} \; = \; \p_{\bf n} V_e \; - \; {\rm i} k A^{\rm \; ret}_{\bf J e \; n}
\label{eqn:s}
\ee
with:
\be
{\rm i} k A^{\rm \; ret}_{\bf J e \; n} \; = \; \f{k^2}{r} p_{\bf n} \; + \; {\rm i} k^3 p_{\bf n} \; + \; {\rm O} \; (kr)^4
\ee
The first term on the right of equation (\ref{eqn:s}) recovers non-retarded eigenvalues, thence equation (\ref{eqn:7}) modifies into:
\be
\lambda^{\rm \; ret}_{\underline{k} s} \; = \; \lambda_{\underline{k} s} \; - \; {\rm i} k \lim_{r,r' \rightarrow \; R} \f{\iint {\rm G}_e \widehat{A}^{\rm \; ret}_{\bf J e \; n} d s d s'}{\iint {\rm G}_i \p_{\bf n} {\rm G}_i d s d s'}
\label{eqn:13-b}
\ee
$\widehat{A}^{\rm \; ret}_{\bf J e \; n}$ expressing the geometric behavior of $A^{\rm \; ret}_{\bf J e \; n}$ (here, the dipole contribution $({\bf n} \cdot {\bf r})_z$) and the denominator descending from equation (\ref{eqn:mode}).
For the spherical symmetry, calculations can be specialized at $R = L/2$ and the final result is (SI 6M):
\be
\lambda^{\rm \; ret}_{\ell = 1} \; = \; - \; 2 \; [1 \; + \; (k R)^2 + \; {\rm i} (k R)^3] \; + \; {\rm O} \; (kr)^4 \;\;\;\;\; ({\rm sphere})
\label{eqn:ret}
\ee
in fair agreement with previous corrections from radiation damping and depolarization,\cite{zeman87} implying $\delta \lambda^{\rm \; ret}_{\ell = 1} = - 3 (k R)^2 - 2 {\rm i} (k R)^3$, and with the second-order term\cite{bohren1998} coming from an analysis of Mie's scattering coefficients, $- \f{12}{5} (k R)^2$.\\
Likewise equation (\ref{eqn:13-b}), eigenvalue corrections for semicubes are:
\be
\delta \lambda^{\rm \; ret}_{\underline{k} b} \; = \; - \; \f{{\rm i} k L}{2 q^2_i }\; \langle \; X_{\underline{k}} \; | \; V_e \; (\widehat{A}^{\rm \; ret}_{\bf J e \; n} \; + \; \delta \widehat{A}^{\rm \; ret}_{\bf \rho e \; n}) \; | \; X_{\underline{k}} \; \rangle
\label{eqn:delta-x}
\ee
in which equation (\ref{eqn:13b}) was employed.
We include one more term in each vector potential, retaining the quadrupole tensor (${\bf Q}$) in the Lorenz-like expression:\cite{jackb}
\be
{\bf A}^{\rm ret}_{\bf J e} = \; - \; \f{{\rm i} k}{r} \left[{\rm e}^{{\rm i} k r} {\bf p} \; + \; \f{y (k r)}{6 r} \; {\bf Q} (\widehat{\bf n}) \right]
\ee
where $y \equiv (1 - {\rm i} k r) {\rm e}^{{\rm i} k r}$ and $\widehat{\bf n} \equiv \widehat{\bf R} \; ({\bf r}' = {\bf 0})$, while the next order in equation (\ref{eqn:n}) returns:
\be
{\bf I}_{_1} \approx \; V_e {\bf r} \; - \; \f{{\bf p}}{r} \; - \; \f{{\bf Q} ({\bf r})}{3 r^3}
\ee
In this way, the second-order correction to semicubic eigenvalues is still real and negative:
\be
- \; {\rm i} k \left( \widehat{A}^{\rm \; ret}_{\bf J e \; n} \; + \; \delta \widehat{A}^{\rm \; ret}_{\bf \rho e \; n} \right) \; = \; 
- \; \f{k^2}{2 r} \left( p_{\bf n} + r r_{\bf n} V_e + \f{\cal Q}{3 r^2}  \right) \; + \; {\rm O} \; (k r)^3
\label{eqn:ret-2}
\ee
with ${\cal Q} \equiv r Q_{\bf n} (\widehat{\bf n}) - Q_{\bf n} ({\bf r}) = 0$ as ${\bf r} = r \widehat{\bf n}$.
Equation (\ref{eqn:ret-2}) is used now in equation (\ref{eqn:delta-x}), and specialized to the particle radius $R^* = R^* ({\rm S})$ of a sphere of equivalent volume.
Let $\lambda_{\underline{k} b}$ to denote the non-retarded equation (\ref{eqn:13}), all contributions to the retarded eigenvalue write, at second order:
\be
\lambda^{\rm \; ret}_{\underline{k} b} \approx \; \lambda_{\underline{k} b} \; + \; \delta \lambda^{\rm \; ret}_{\ell = 1} \; - \; \tfrac{1}{2} \Delta \lambdabar^2 \left( \f{k R^*}{\lambdabar_{_0}} \right)^2 \langle \; X_{\underline{k}} \; | \; u_{\ell} \; (1 \; + \; u_{\ell}) \; ({\bf n} \cdot \widehat{\bf r})_z \; | \; X_{\underline{k}} \; \rangle
\label{eqn:ret-3}
\ee
where it was taken into account equation (\ref{eqn:14}), $\sqrt{\Delta \lambdabar^2} \approx (q_{\underline{k} b} / q_i) \lambdabar_{_0}$, and that geometric dipole-like contributions stemming from $p_{\bf n}/r$ and $r_{\bf n} V_e$ are the same as, from equation (\ref{eqn:10b}):
\be
u_{\ell} \; ({\bf r}) \; = \; \sum^{2 \ell}_{2 {\rm N} \; \ni \; \ell' \; > \; 1} \o{r^{\ell'}} \f{{\rm L}_{\ell \ell'}}{r^{\ell'}}
\label{eqn:u-l}
\ee
To work out equation (\ref{eqn:ret-3}), remind that $P$ terms do not split in fields of cubic symmetry.
Therefore, the dominant correction for $\ell = 1$ was taken to be spherical (equation \ref{eqn:ret}) while, correspondingly, the first addendum in equation (\ref{eqn:u-l}) is evaluated for $\ell = 2$.
The main calculations for the average in equation (\ref{eqn:ret-3}) are still gathered in SI (6M), and return, for $\underline{k} = \{ 2 0 \}$:
\be
\tfrac{1}{2} \langle \; X_{\underline{k}} \; | \; u_{\ell} \; (1 \; + \; u_{\ell}) \; ({\bf n} \cdot \widehat{\bf r})_z \; | \; X_{\underline{k}} \; \rangle \; = \; \tfrac{1}{66} \left( 23 \; + \; \tfrac{2527}{39} \tfrac{\overline{r^4}}{{R^*}^4} \right) \tfrac{\overline{r^4}}{{R^*}^4}
\label{eqn:lbo}
\ee
implying, in the same fashion of equation (\ref{eqn:14}): 
\be
\Delta \lambdabar^2_{ret} \; = \; \left[ \tfrac{1}{66} \tfrac{\overline{r^4}}{{R^*}^4} \left( 23 \; + \; \tfrac{2527}{39} \tfrac{\overline{r^4}}{{R^*}^4} \right) \Delta \lambdabar^2 \; + \; 2 \lambdabar^2_{_0} \right] (k R^*)^2 \; + \; {\rm O} (k R^*)^3
\label{eqn:last}
\ee
with $\Delta \lambdabar^2_{\rm \; ret} = {\lambdabar^{\rm \; ret}}^2 - \lambdabar^2$ now being defined as the second-order difference between the squares of retarded and non-retarded wavelengths. 
To use the last relation only requires to recover the non-retarded best fit and the expression $R^* ({\rm S}) = (29.92 \; {\rm S} + 11.80)$ nm (SI 5F).
In Figure (\ref{fig:ret}) are two applications, one to the retarded-geometrically corrected data (sc), the second to the experimental measurements (ex).
The latter is affected by an excess charge injection term ($\xi$), increasing the plasma frequency by a factor of $\sqrt{1 + \xi}$.
The wavelength $\lambdabar_{_0}$ thus was lowered accordingly, based on the exact correction $\xi ({\rm S}) = 0.213 \; {\rm S} - 0.041$ best fitting the numerical computations (SI 5F).
As to the average spatial scale, which in the original CFT roughly identifies Bohr's radius,\cite{figgis00} $\overline{r^4} \sim a^4_{_0}$, here is expected to range in the nm scale (some lattice constants).
Best fits, in fact, return $\overline{r^4} \sim 1$ nm$^4$ in both circumstances (see caption to Figure \ref{fig:ret}). 

\begin{figure}[h]
	\centering
	\includegraphics[scale=0.0415]{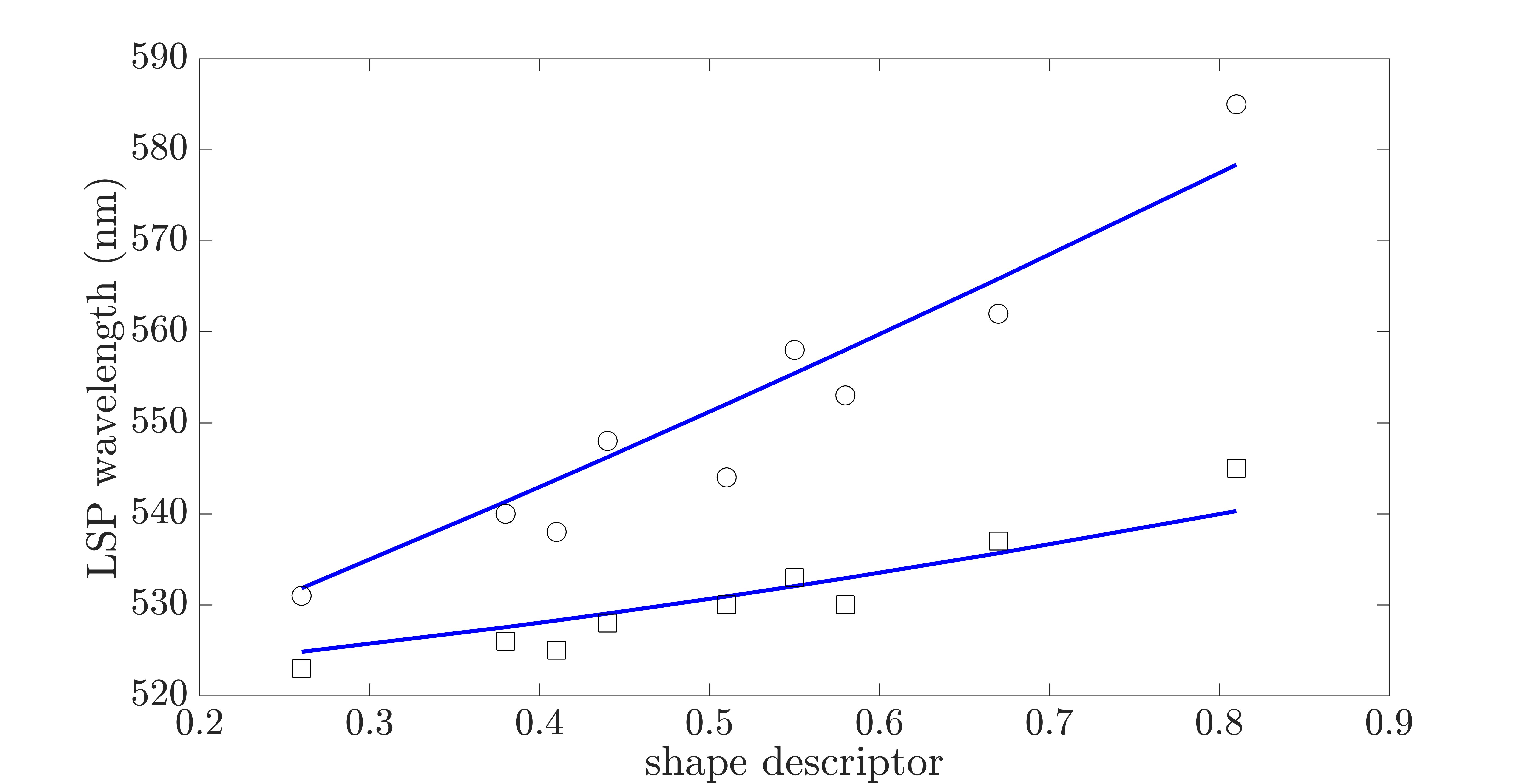}
	\caption{Retarded LSP wavelength $\lambdabar^{\rm \; ret}$ vs shape descriptor ${\rm S}$ (points), based on the interpolation (blue line) in Figure (\ref{fig:lasfig}) and corrected by equation (\ref{eqn:last}). Upper line interprets the retarded and geometrically corrected data, $\lambdabar \; ({\rm sc})$ in Table \ref{tbl:tab2} (SI 2), with $k = 11.4$ $\mu$m$^{-1}$.
	Lower line fits the experimental data, $\lambdabar \; ({\rm ex})$, where $k = 11.8$ $\mu$m$^{-1}$ and $\lambdabar_{_0}$ was excess-charge corrected (see text).
	Best fits yield $\overline{r^4} = 0.98$ ($R^2 = 0.91$) and $0.99$ nm$^4$ ($R^2 = 0.86$), respectively.}
	\label{fig:ret}
\end{figure}

\noindent {\rm \bf Eigenvalues and hybridization in cubes}. The accuracy best fits were conducted with (average determination coefficient $= 0.99$) is encouraging to get reliable extrapolations of cube eigenvalues ($\lambda_{\underline{k} b} \rightarrow \lambda_{\underline{k} c}$ for ${\rm S} \rightarrow 1^-$), as they can suffer from numerical instabilities due to wedge- and vertex-like effects.\\
Values of $\lambda_{\underline{k} c}$ are inferred from polynomial coefficients in SI (6L) and classified in Table (\ref{tbl:cube}) in terms of the spherical $\ell$, the degeneracy of $m$ states and their symmetry character.
These trends agree with former studies on resonating cubes,\cite{langbein75,fuc75,hohen05} and seem to suggest hybridization of a number of $\ell$-bands (e.g. see $\ell = 5$).

\begin{table}[h]
\small
  \caption{Cube eigenvalues extrapolated from the first five spherical $\ell$ values.}
  \label{tbl:cube}
  \begin{tabular*}{0.48\textwidth}{@{\extracolsep{\fill}}|llll|llll|}
    \hline
        $\ell$ & deg & sym &  $-\lambda_{\underline{k} c}$ & $\ell$ & deg & sym &  $-\lambda_{\underline{k} c}$ \\
   	\hline
1 & 3 & z & 4.00 & 3 & 3 & xy & 2.23 \\
2 & 3 & xy & 3.97 & 4 & 3 & xyz, x$^2$ - y$^2$ & 2.17 \\
3 & 1 & xyz & 3.96 & 5 & 3 & xy, x$^2$ - y$^2$ & 2.16 \\
4 & 1 & xyz & 3.37 & 5 & 2 & z & 2.12 \\
3 & 3 & z & 2.79 & 5 & 1 & z & 2.12 \\
4 & 3 & xy & 2.74 & 4 & 2 & xy & 1.96 \\
2 & 2 & x$^2$ - y$^2$ & 2.36 & 5 & 1 & xy & 1.90 \\
5 & 3 & z, x$^2$ - y$^2$ & 2.36 & 5 & 2 & z, x$^2$ - y$^2$ & 1.88 \\
\hline
  \end{tabular*}
\end{table}


\noindent {\rm \bf Conclusive remarks}. A sphere-to-cube transition in plasmonic Au nanoparticles is experimentally and numerically described in detail.
An analysis of spontaneous electric oscillations (normal modes) is carried out in light of crystal (or ligand) field theory (CFT), allowing a reinterpretation of geometric eigenmodes and resonant plasmon wavelength.
This relationship is promising to better understand the role of shape in plasmonics and nanophotonics, continuing a cross-fertilization with theoretical chemistry started almost two decades ago with particle hybridization.\cite{pro03,san12}
Here, a plasmon splitting concept is inferred from the electromagnetic analog of CFT, both in a retarded and non-retarded picture.\\ 
From a broader perspective, the theoretical insights here provided can have important implications in assessing the exact nanocrystal topology (in this case, from sphere to cube) by merely optical means.
In addition, the well-resolved relation between the maximum of LSP resonance and shape descriptor ({\rm S}) can serve like a useful tool in plasmon-based biosensing, as it either relies on refractive index changes\cite{lee_refractive_2011,chen08} or nanocrystal shape transitions in response to a certain degree of oxidative stress in a given medium.\cite{coronado-puchau_enzymatic_2013}   


\newpage

\section*{Acknowledgements}

This work was supported by the Ministry of Environment and Energy, the Ministry of Science and Education, the Environmental Protection and Energy Efficiency Fund, and the Croatian Science Foundation (HrZZ) under the project "Plasmonic Alternative Materials for Solar Energy Conversion" (PKP-2016-06-4469) in the total amount of 1 074 000 HRK.\\
S.A.M. thanks DIPC (Donostia-San Sebastian, Spain) and IRB (Zagreb - Croatia) for kind hospitality and financial support.\\
Ana S\'anchez Iglesias is kindly acknowledged for her help during the synthesis of nanoparticles. 

\section*{Author Contributions}

The authors equally contribute to this work (A.S.I and M.G. by the experimental part, J.S.P. by the computational session, S.A.M. by the theoretical formulation).

\section*{Additional Information}

\subsection*{Supporting Information (SI)}

A supporting information file accompanies this paper.

\subsection*{Competing Interests}

Authors declare no competing interests.

\newpage\phantom{blabla}

\section*{\rm \bf \underline{\LARGE{Supporting Information (SI)}}}

\renewcommand{\thefigure}{S\arabic{figure}}
\setcounter{figure}{0}

\section*{\rm \bf 1. Methods}

\subsection*{\rm \bf A. Experimental}

\textbf{Chemicals:} Gold (III) chloride trihydrate (\ce{HAuCl4 * 3H2O}), sodium borohydride (\ce{NaBH4}) cetyltrimethylammonium bromide (CTAB), ascorbic acid (AA), cetyltrimethylammonium chloride (CTAC, 25 wt.\% in water), benzyldimethylhexadecylammonium chloride (BDAC), sodium hypochlorite, and sodium bromide (\ce{NaBr}) were purchased from Sigma-Aldrich and used without further purification. Milli-Q water was used in all experiments. \\
\textbf{Synthesis of 27 nm single-crystaline gold seeds:}\cite{zheng_successive_2014} Gold seeds nanoparticles were obtained through two-step overgrowth of initial single crystalline seeds of 1-2 nm. Typically, initial seeds were prepared by reduction of \ce{HAuCl4} (5 mL, 0.25 mM) with freshly prepared \ce{NaBH4} (0.3 mL, 10 mM) in aqueous CTAB solution (100 mM). The mixture was left undisturbed at 27 \textdegree C for 0.5 h to ensure complete decomposition of \ce{NaBH4}. An aliquot of seed solution (0.11 mL) was added to a growth solution containing CTAC (20 mL, 200 mM), \ce{HAuCl4} (20 mL, 0.5 mM) and AA (15 mL, 100 mM). The mixture was left undisturbed at 27 \textdegree C for 30 min. The solution was centrifuged (1 h, 14000 rpm) to remove excess of reagents and redispersed in water to obtain a final concentration of gold equal to 3 mM. The resulting solution of gold nanospheres (10 nm) was used as seeds in the second overgrowth to obtain 27 nm nanoparticles. A solution of gold nanospheres of 10 nm in diameter (0.285 mL, 3 mM) was added under vigorous stirring to an aqueous growth solution of BDAC (50 mL, 100 mM), \ce{HAuCl4} (0.25 mL, 50 mM) and AA (0.25 mL, 100 mM) at 40 \textdegree C. The mixture was left undisturbed at 30 \textdegree C for 30 min. The solution was centrifuged twice (6500 rpm, 30 min) to remove excess of reagents, and redispersed in water to obtain a final concentration of gold equal to 7.5 mM. \\
\textbf{Synthesis of nanocubes:} To three solutions containing CTAC (5 mL, 15 mM) and Au seeds of 27 nm (0.122 mL, 0.46 mL, 0.24 mL, 7.5 mM, for 40, 50 and 60 nm cubes, respectively) was added NaBr (0.5 mL, 10 mM), AA (0.0188 mL, 100 mM) and \ce{HAuCl4} (0.025 mL, 50 mM). The mixtures were left under stirring for 30 min at room temperature. \\
\textbf{Oxidative etching:} To the as-prepared solutions of gold nanocubes was added a solution of sodium hipochlorite (0.018 mL, 1 vol\%). The mixtures were left for 5 min at room temperature under gentle stirring. Then, the solutions were centrifugated and redispersed in CTAB (100 mM), followed by two-cycle centrifugation and redispersion in water. \\  
\textbf{Image analysis}: To obtain the (uncorrected) values of length and circularity for each sample, a Transmission Electron Microscopy (TEM) characterization was carried out followed by image analysis using the open-source Fiji software endowed with BioVoxxel plugin.
For all samples, raw images (*.dm3) at $20000$x magnification were used.
The original grey-scale image was converted to a binary one by a conversion into 8-bit and implementing a color-scale threshold with minimal method.
A median filter with radius $= 2$ (dimensionless units) then was applied to smooth each particle edges, while, to separate connected units, a watershed algorithm was deployed.
Values of length and circularity came from the Extended Particles Analysis (BioVoxxel), implemented by constraining the surface area to range within $500-6000$ \ce{nm^2}.
Such a workflow allowed for the characterization over up to $400$-particle sample from a single image.

\subsection*{\rm \bf B. Numerical}

Particle surfaces were discretized by allowing a maximum surface area element of $3$ nm$^2$ and refining every domain no more than $10$ nm away from vertices and edges.
To increase resolution in these areas then was mandatory to guarantee an accurate description of charge accumulation when plasmon resonance gets excited.\cite{hohenester2012mnpbem}
Convergence of numerical simulations were excellent, results being stable upon refining the discretization any further.
The number of surface elements was $\approx 104$, the double mirror symmetry of particles substantially reducing the computational costs (by a factor $\sim 6$).\\
In the MNPBEM code,\cite{hohenester2012mnpbem}, scalar and vector potentials are expressed via boundary integrals (i.e. sum over boundary elements) of surface charge and current distributions coming from the boundary conditions of Maxwell's equations.
The electromagnetic field then is calculated at any spatial point, and the relevant optical quantities (extinction, field enhancement, scattering and absorption cross sections) are finally computed.
Such an implementation was used with full Maxwell's equations, the quasistatic approximation and eigenmode expansion.
Optical spectra follow from literature values of the dielectric functions of Au single crystals\cite{olmon2012optical} and water.\cite{Hale:73}

\section*{\rm \bf 2. Statistical Analysis of Geometric Nanoparticle Features}

Information on the shape descriptor and radius of spherical caps stems from an ad-hoc analysis of circularity and size distributions determined from each sample (Experimental Section - Image analysis).
In the next subsections, we explain how all the geometrical data in Table (\ref{tbl:tab2}) were inferred.

\begin{table*}[h]
  \caption{Geometric properties of nanoparticle ensembles. ${\rm S}$ (unc) is the shape descriptor got straight away from the image analysis of ${\rm C}$. ${\rm S}$ comes from equation (\ref{eqn:b14}).}
  \label{tbl:tab2}
  \begin{tabular*}{\textwidth}{@{\extracolsep{\fill}}lllll}
    \hline
        $\langle$ {\em L} $\rangle$ \; [nm] & {\rm C} & {\rm S} (unc) & {\rm S} & $\langle$ {\em R} $\rangle$ \; [nm]\\
		\hline
		 38.1 $\pm$ 2.2 & 0.87 $\pm$ 0.02 & 0.72 $\pm$ 0.06 & 0.51 $\pm$ 0.08 & 5.3 $\pm$ 1.2 \\
		 35.8 $\pm$ 1.3 & 0.89 $\pm$ 0.02 & 0.66 $\pm$ 0.06 & 0.41 $\pm$ 0.08 & 6.0 $\pm$ 1.1 \\
		 34.3 $\pm$ 1.1 & 0.91 $\pm$ 0.01 & 0.61 $\pm$ 0.03 & 0.26 $\pm$ 0.06 & 6.7 $\pm$ 0.6 \\
		 49.5 $\pm$ 3.4 & 0.84 $\pm$ 0.03 & 0.82 $\pm$ 0.09 & 0.67 $\pm$ 0.10 & 4.5 $\pm$ 2.3 \\
		 46.0 $\pm$ 1.9 & 0.87 $\pm$ 0.02 & 0.75 $\pm$ 0.06 & 0.58 $\pm$ 0.07 & 5.8 $\pm$ 1.4 \\
		 43.5 $\pm$ 1.0 & 0.90 $\pm$ 0.01 & 0.63 $\pm$ 0.03 & 0.38 $\pm$ 0.04 & 8.1 $\pm$ 0.6 \\
		 60.2 $\pm$ 3.6 & 0.81 $\pm$ 0.02 & 0.92 $\pm$ 0.07 & 0.81 $\pm$ 0.08 & 2.5 $\pm$ 2.2 \\
		 54.9 $\pm$ 1.9 & 0.88 $\pm$ 0.02 & 0.70 $\pm$ 0.07 & 0.55 $\pm$ 0.08 & 8.2 $\pm$ 1.9 \\
		 50.8 $\pm$ 1.2 & 0.90 $\pm$ 0.01 & 0.64 $\pm$ 0.05 & 0.44 $\pm$ 0.06 & 9.1 $\pm$ 1.2 \\
    \hline
  \end{tabular*}
\end{table*}

\subsection*{\rm \bf A. Shape Descriptor from Circularity Values}

\renewcommand{\theequation}{A.\arabic{equation}}
\setcounter{equation}{0}

Circularity is a geometrical descriptor for a surface in two dimensions, which is proportional to the ratio between area ($A$) and the square of perimeter ($P$):
\be
{\rm C} = 4 \pi \f{A}{P^2}
\label{eqn:a1}
\ee
The angle $4 \pi$ is introduced, for a perfect circle, to get the unit value ${\rm C}_\circ = 1$, implying a perfect square to take ${\rm C}_q = \tfrac{\pi}{4}$.\\
Given a distribution of circularity values, the first step is deriving the average shape descriptor from it.
For a semicube, the former is defined by the radius of spherical caps at the corners and length of rectilinear edges:
\be
{\rm S} \; = \; 1 - \f{2 R}{L} 
\label{eqn:a2}
\ee
It follows that the algeabric equation:
\be
c_{_2} {\rm S}^2 + \; c_{_1} {\rm S} \; + \; c_{_0} = 0
\label{eqn:a3}
\ee
where the functions $c_i = c_i ({\rm C})$ may be expressed as:
\begin{align}
c_{_0} & = \tfrac{1}{{\rm C}_q} ({\rm C} - {\rm C}_\circ)
\\
c_{_1} & = 2 ({\rm C}_\circ - {\rm C}_q) ({\rm C} - {\rm C}_\circ)
\\
c_{_2} & = ({\rm C}_q - {\rm C}_\circ) ({\rm C} - {\rm C}_\circ) + (\tfrac{1}{{\rm C}_q} - {\rm C}_\circ) {\rm C}
\label{eqn:a6}
\end{align}
identify the sought application ${\rm S} =  {\rm S} ({\rm C})$, since ${\rm S} ({\rm C}_\circ) = 0$ (sphere), ${\rm S} ({\rm C}_q) = 1$ (cube) in the two ideal cases $R = 0, \tfrac{L}{2}$.\\
Equations (A4-A6) map a 2d description to 3d, increasing the root multiplicity by one unit.
For this reason, any perturbation $\delta {\rm C}$ will generate a larger $\delta {\rm S}$, one of the two ${\rm S}$ roots being negative and thus to be discarded.
Standard deviations, $\sigma_{_C}$ and $\sigma_{_S}$, are related instead by letting statistical and systematic errors propagate through equation (\ref{eqn:a3}):
\be
\sigma_{_S} = \; \left| \f{d {\rm S}}{d {\rm C}} \right|_{\langle {\rm C} \rangle} \sigma_{_C}
\label{eqn:a7}
\ee
Working every perturbation out, the derivative writes:
\be
\left( \f{d {\rm S}}{d {\rm C}} \right) \; = \; \f{{\rm S}}{2 \langle c \rangle_{_0} + \langle c \rangle_{_1} {\rm S}} \left( \f{\langle c \rangle_{_0} + \langle c \rangle_{_1} {\rm S}}{{\rm C} - {\rm C}_\circ} + \f{\langle c \rangle_{_2} - {\rm C}_\circ + {\rm C}_q}{{\rm C}}  
\right)
\label{eqn:a8}
\ee
where, for notational simplicity, ${\rm C}$ and ${\rm S}$ here denote the {\em average values} of circularity and shape descriptor.

\subsection*{\rm \bf B. Statistical Correction to Circularity}

\renewcommand{\theequation}{B.\arabic{equation}}
\setcounter{equation}{0}

The finite size of pixels, here $\sigma_p \approx 0.5$ nm, represent itself a systematic uncertainty affecting circularity.
To correct the former statistics, consider a random and independent error process, keeping the variance unchanged but altering the mean value as:
\be
\f{\delta {\rm C}}{{\rm C}} = \sqrt{ \left(\f{\delta A}{\langle A \rangle} \right)^2 + \; 4 \left( \f{\delta P}{\langle P \rangle} \right)^2}
\label{eqn:b9}
\ee
We also suppose, as a consequence of chemical etching, the contribution from the particle curvature to be dominant, i.e. let:
\be
A \; = \; \pi R^2 + l^2 + 4 l R \;,\;\;
P \; = 4 l + 2 \pi R
\label{eqn:b10}
\ee
then:
\be
\delta A \; \approx \; 2 (\pi R + 2 l) \sigma_p \;,\;\;
\delta P \; \approx \; 2 \pi \sigma_p
\label{eqn:b11}
\ee
with $\delta R \approx \sigma_p$ and $\f{\langle l \rangle}{\langle L \rangle} \equiv {\rm S}$.
To benefit from a reasonable formal approximation, as $2 < \pi < 4$, we take for the quantity $2 l + \pi R$ the arithmetic average of the extremes $2 L$ and $L + l$, while $\pi R^2 + l^2 + 4 l R \approx L^2$, thence:
\begin{align}
\f{\delta A}{\langle A \rangle} & \approx \; \left(3 + {\rm S} \right) \f{\sigma_p}{\langle L \rangle} \\
\f{\delta P}{\langle P \rangle} & \approx \; \left( \f{2 \pi}{3 + {\rm S}} \right) \f{\sigma_p}{\langle L \rangle}
\label{eqn:b13}
\end{align}
and the final relation, expressing a systematic correction to the mean circularity, is:
\be
\delta {\rm C} \; \approx \; \f{\sigma_p {\rm C}}{ \langle L \rangle (3 + {\rm S})} \sqrt{(3 + {\rm S})^4 + 16 \pi^2}
\label{eqn:b14}
\ee
As the shape descriptor in equation (\ref{eqn:a3}) is to be corrected by equation (\ref{eqn:b14}), one is left with a coupled equation systems for ${\rm S} = {\rm S} ({\rm C})$, which was solved to the second digit.

\subsection*{\rm \bf C. Statistics of Radii}

\renewcommand{\theequation}{C.\arabic{equation}}
\setcounter{equation}{0}

To get average value and standard deviation of particle radii, consider two normal distributions as:
\be
L \; \sim \; {\rm N} (\langle L \rangle, \sigma_{_L}) \;,\;\;
R \; \sim \; {\rm N} (\langle R \rangle, \sigma_{_R})
\label{eqn:c15}
\ee  
the first of which being experimentally accessible.
While $\langle R \rangle = \f{1}{2} (1 - {\rm S}) \langle L \rangle$, deriving the unknown $\sigma_{_R}$ requires to introduce a product distribution\cite{dale79} for two aleatory variables, $A B$, for which:
\be
\sigma^2_{a b} \; = \; \sigma^2_a \sigma^2_b \; + \; \langle A \rangle^2 \sigma^2_b \; + \; \langle B \rangle^2 \sigma^2_a
\label{eqn:c16}
\ee
Let e.g. $A \equiv L$ and $B \equiv R/L$, one obtains:
\be
\langle B \rangle = \tfrac{1}{2} (1 - {\rm S}) \;,\;\;
\sigma_{_B} = \tfrac{\sigma_{_S}}{2}
\label{eqn:c17}
\ee
and, by exploiting equation (\ref{eqn:c16}), it turns out:
\be
\sigma_{_R} = \f{\sigma_{_L} \sigma_{_S}}{2} \sqrt{1 + \left (\tfrac{\langle L \rangle}{\sigma_{_L}} \right)^2 + \left( \tfrac{1 - {\rm S}}{\sigma_{_S}} \right)^2}
\label{eqn:c18}
\ee
where the units of measure are consistent as $\sigma_{_S}$ is dimensionless.

\section*{\rm \bf 3. Geometric Eigenmodes}

We report here the shape-like evolution of some geometric modes for ${\rm S} = 0, \tfrac{1}{3}, \tfrac{2}{3}, 1$.
In Figure (\ref{fig:eigenmode_l1}) are the $3$ dipole modes ($\ell = 1$), while Figure (\ref{fig:eigenmode_l2A}) and (\ref{fig:eigenmode_l2B}) depict the $3 + 2$ modes of the quadrupole ($\ell = 2$).
The $1 + 3 + 3$ octupole modes ($\ell = 3$) are in Figure (\ref{fig:eigenmode_l3A}), (\ref{fig:eigenmode_l3B}) and (\ref{fig:eigenmode_l3C}), respectively.

\begin{figure}[h]
	\centering
	\includegraphics[scale=0.2]{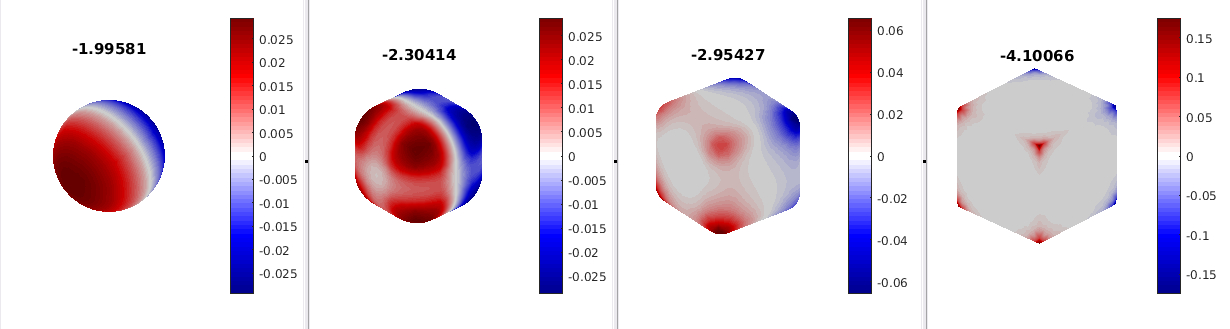}
	\caption{Shape-like evolution of dipole modes ($\ell = 1$).}
	\label{fig:eigenmode_l1}
\end{figure}	

\begin{figure}[h]
	\centering
	\includegraphics[scale=0.2]{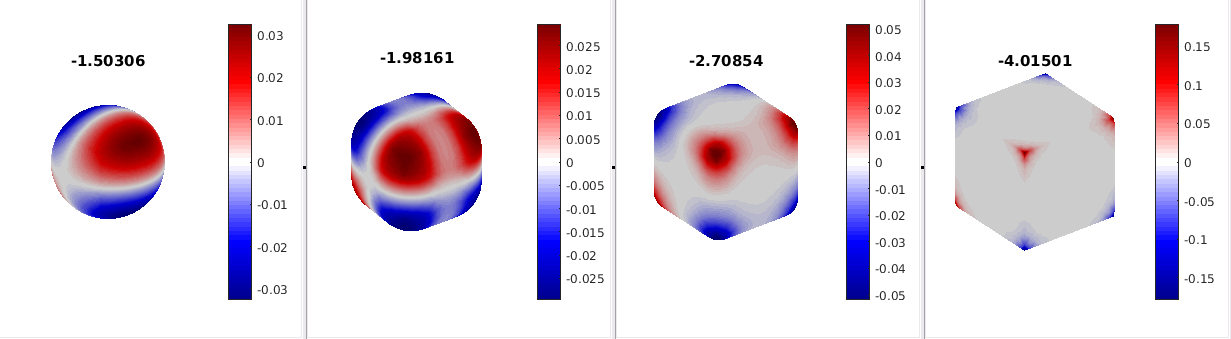}
	\caption{Shape-like evolution of quadrupole corner modes ($\ell = 2$, $T_{2g}$).}
	\label{fig:eigenmode_l2A}
\end{figure}

\begin{figure}[h]
	\centering
	\includegraphics[scale=0.2]{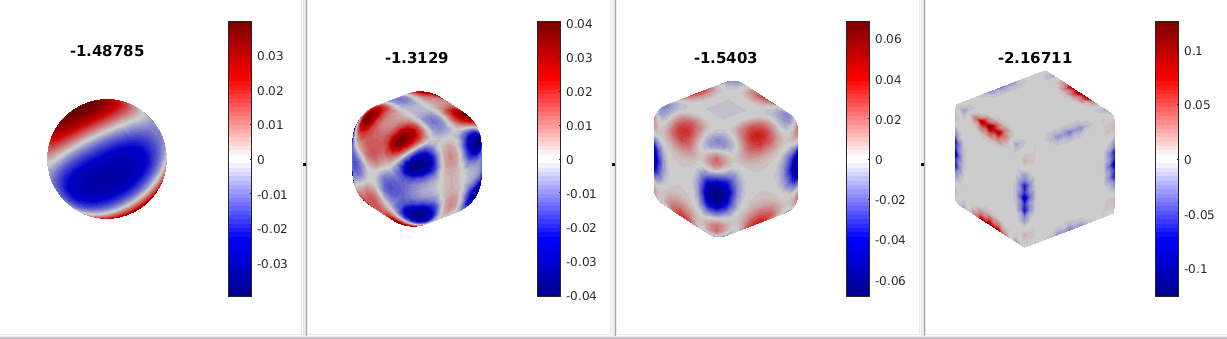}
	\caption{Shape-like evolution of quadrupole edge modes ($\ell = 2$, $E_{g}$).}
	\label{fig:eigenmode_l2B}
\end{figure}

\begin{figure}[h]
	\centering
	\includegraphics[scale=0.2]{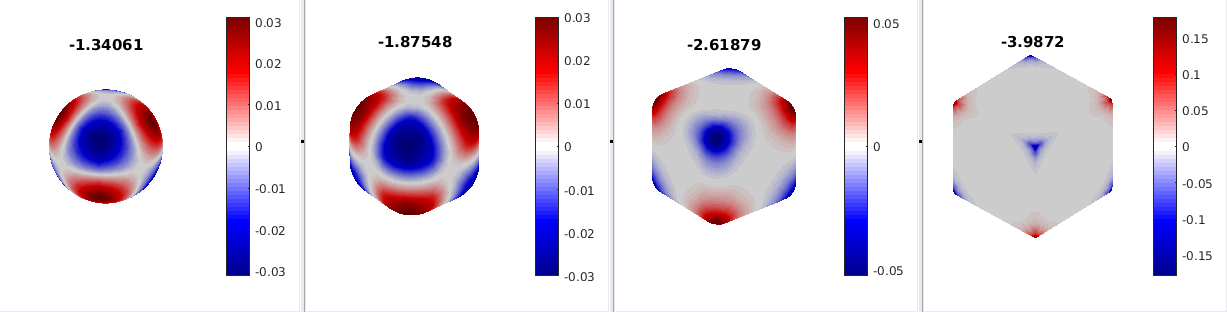}
	\caption{Shape-like evolution of octupole corner modes ($\ell = 3$, $A_{g}$).}
	\label{fig:eigenmode_l3A}
\end{figure}

\begin{figure}[h]
	\centering
	\includegraphics[scale=0.2]{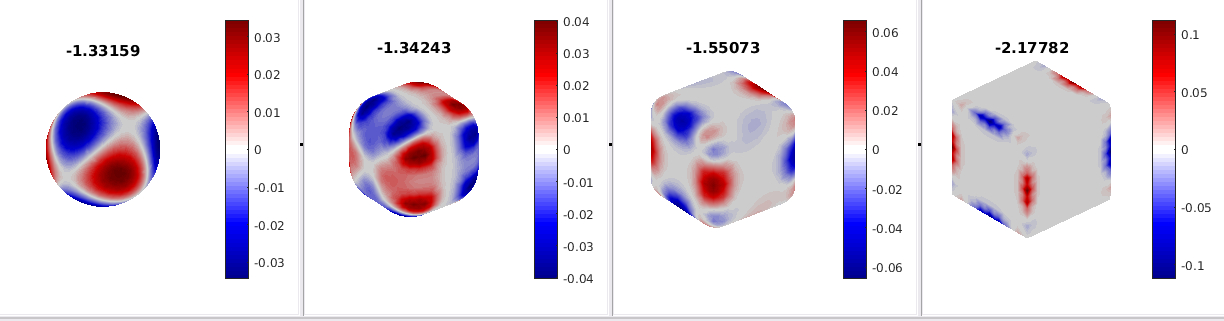}
	\caption{Evolution of octupole edge modes ($\ell = 3$, $T_{2 g}$).}
	\label{fig:eigenmode_l3B}
\end{figure}

\begin{figure}[h]
	\centering
	\includegraphics[scale=0.2]{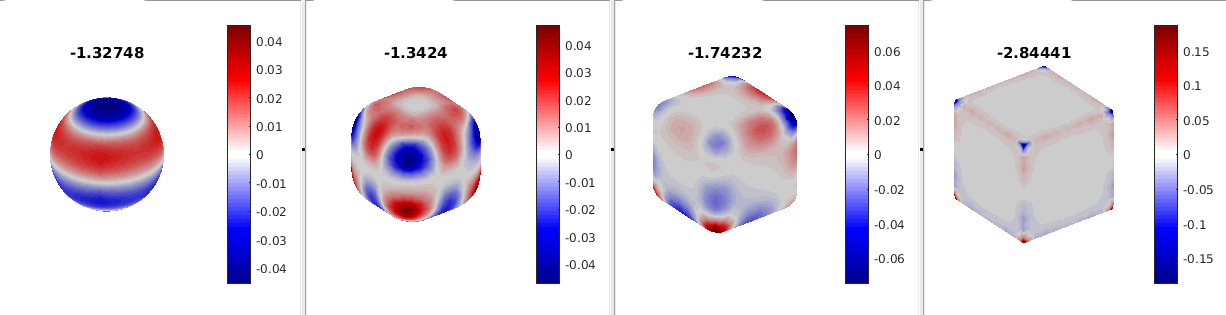}
	\caption{Evolution of octupole corner-edge modes ($\ell = 3$, $T_{1g}$).}
	\label{fig:eigenmode_l3C}
\end{figure}

\section*{\rm \bf 4. Multipole contributions to LSP spectra in Ag particles}

Figures (\ref{fig:ag}, \ref{fig:ag1}, \ref{fig:ag2}, \ref{fig:ag3}) illustrate an example of the extinction spectra for Ag semicubes.
It is seen that, with respect to Au, Ag nanoparticles display three peaks, involving higher-order contributions (up to $\ell = 3$).

\begin{figure}[h]
	\centering
	\includegraphics[width=.5\textwidth,right]{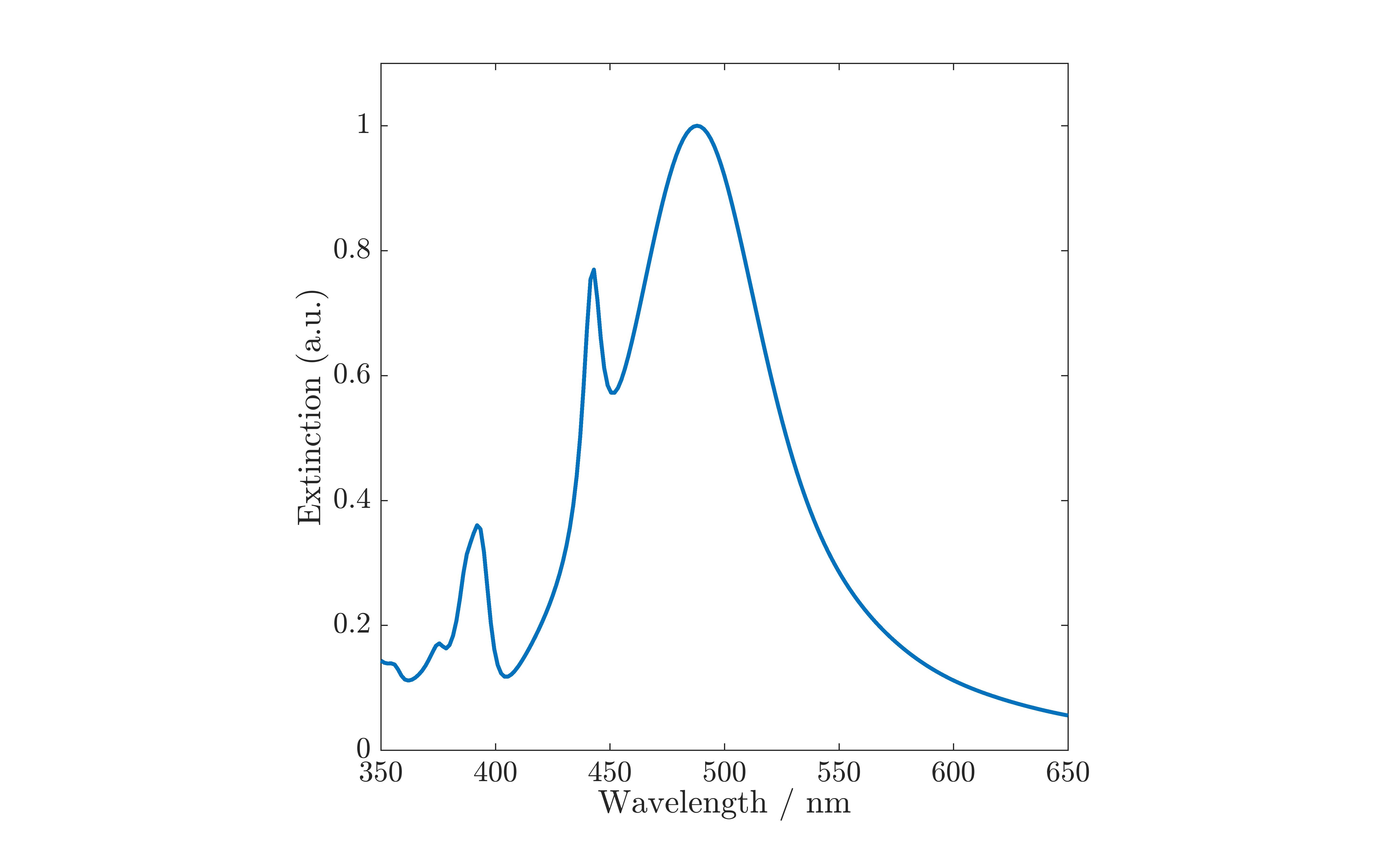}
	\caption{Extinction spectrum in Ag semicube with ${\rm S} = 0.81$.}
	\label{fig:ag}
\end{figure}	

\begin{figure}[h]
	\centering
	\includegraphics[width=.31\textwidth]{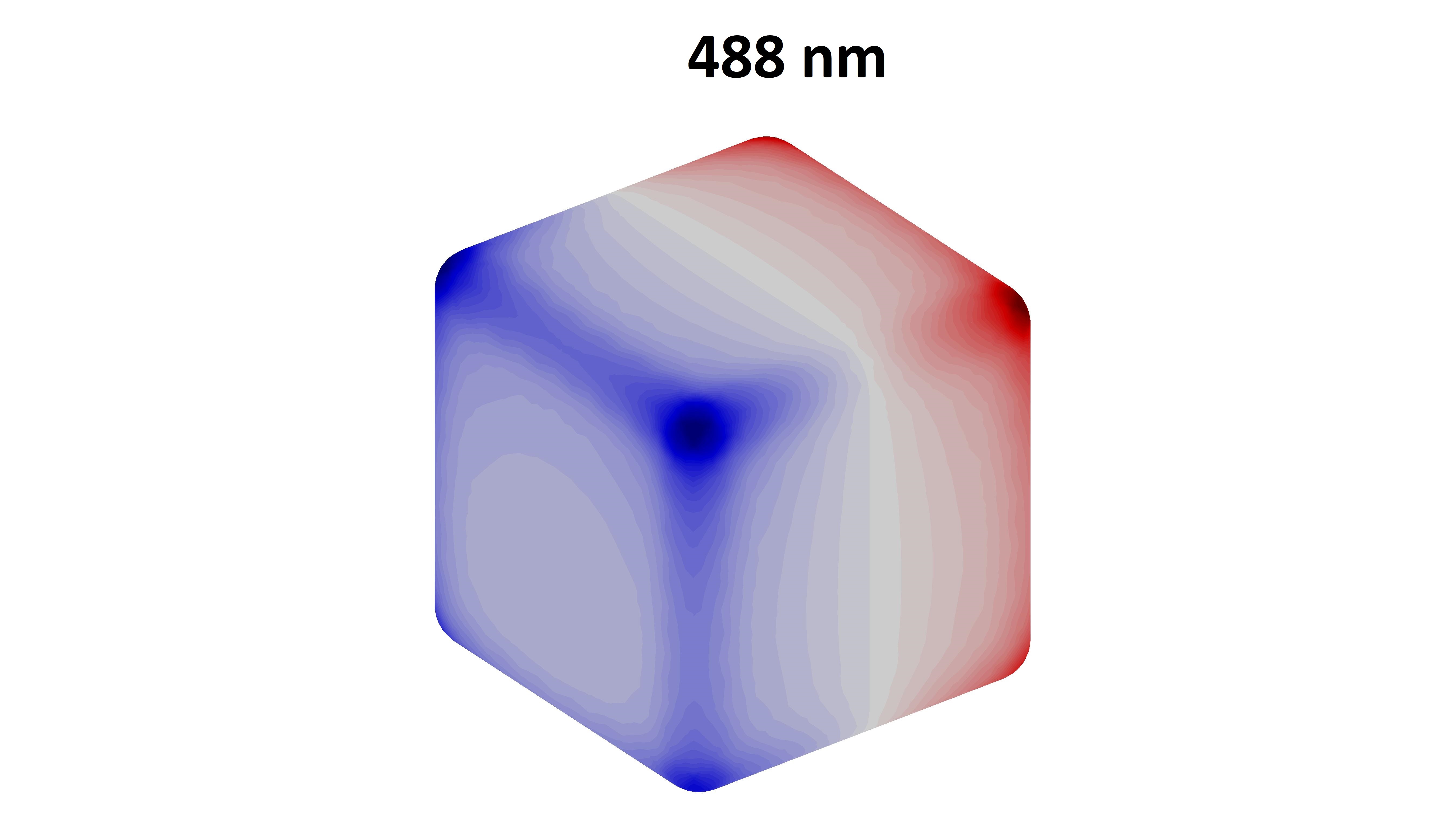}
	\caption{Surface charge distributions in Ag nanocubes (${\rm S} = 0.81$) corresponding to the peak $\ell = 1$ in Figure (\ref{fig:ag}).}
	\label{fig:ag1}
\end{figure}	

\begin{figure}[h]
	\centering
	\includegraphics[width=.31\textwidth]{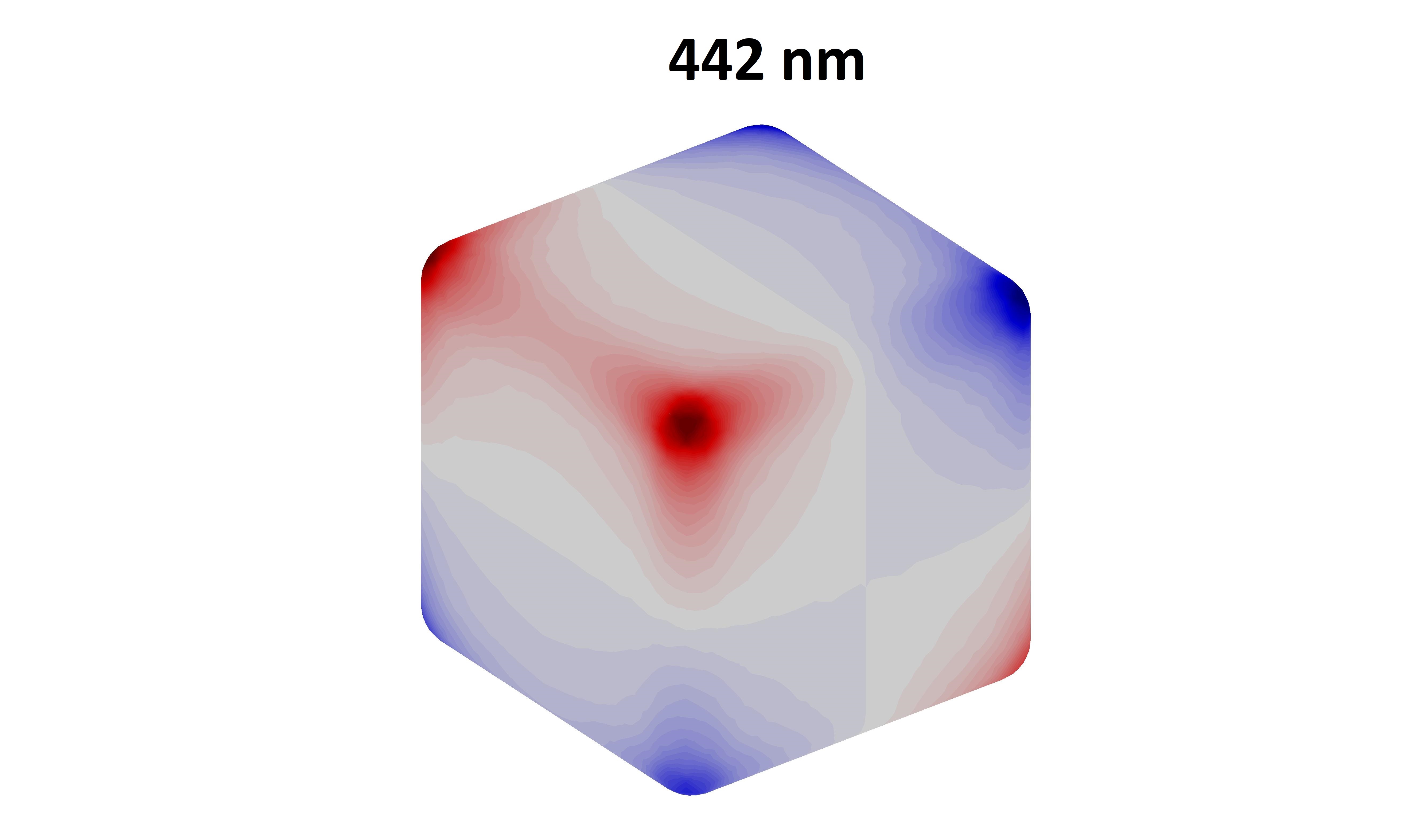}
	\caption{Surface charge distributions in Ag nanocubes (${\rm S} = 0.81$) corresponding to the peak $\ell = 2$ in Figure (\ref{fig:ag}).}
	\label{fig:ag2}
\end{figure}

\begin{figure}[h]
	\centering
	\includegraphics[width=.31\textwidth]{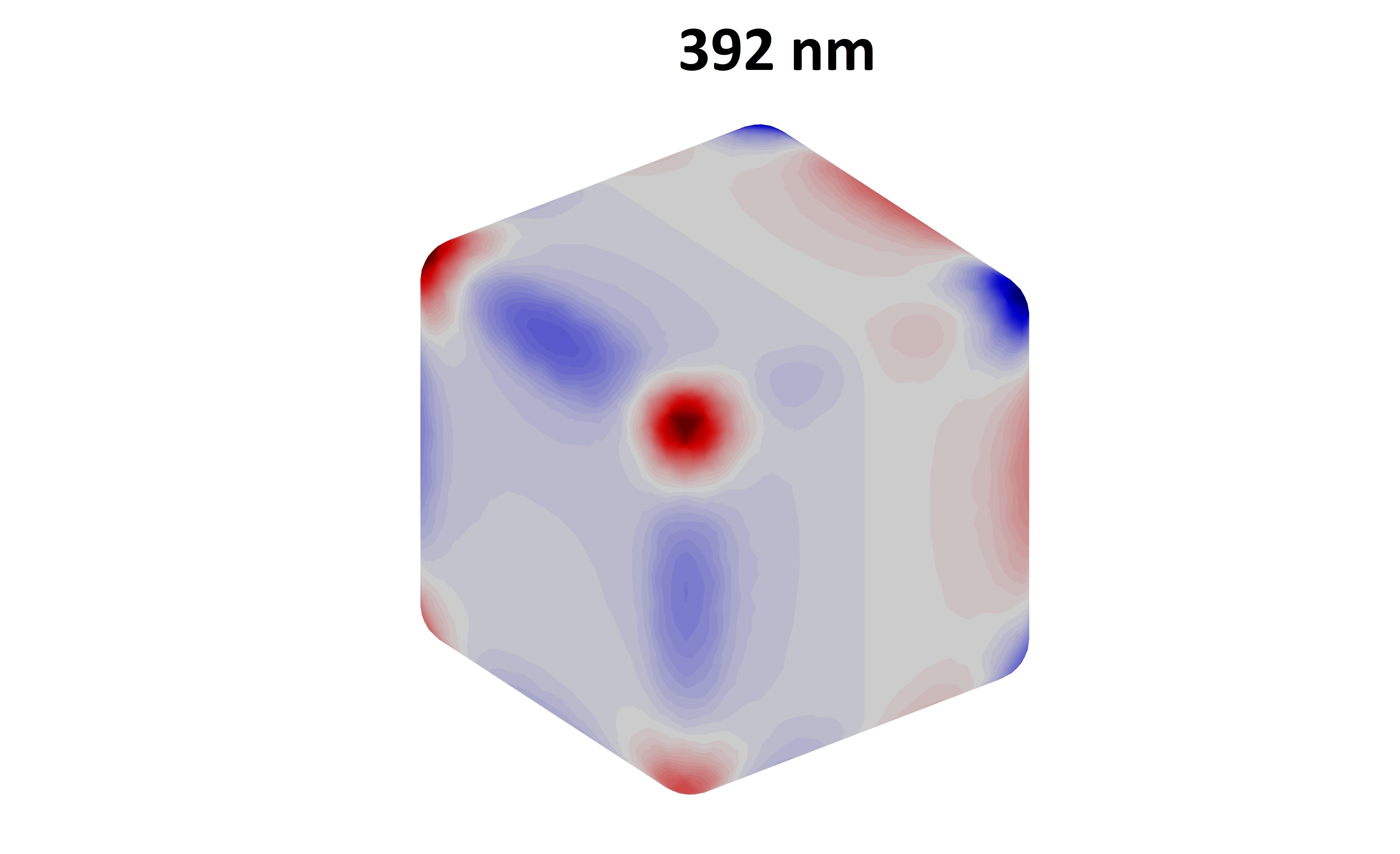}
	\caption{Surface charge distributions in Ag nanocubes (${\rm S} = 0.81$) corresponding to the peak $\ell = 3$ in Figure (\ref{fig:ag}).}
	\label{fig:ag3}
\end{figure}


\section*{\rm \bf 5. Inquiring Anomalous Redshift Sources}

The former statistical correction turned out to be satisfactory, as plasmon wavelengths calculated accordingly better simulate experimental values.
However, there is still a systematic red-shift, specially affecting the largest and least circular particles.
On inquiring three relevant phenomenologies, an explanation for this undesired effect will be looked for in the next subsections.

\subsection*{\rm \bf D. Dielectric Constant Reduction upon Ion Concentration}

\renewcommand{\theequation}{D.\arabic{equation}}
\setcounter{equation}{0}

Electrolytes can influence the dielectric response of environments such as a liquid solution.
Although this effect can be rather small, the high precision targeted by the simulations should avoid refractive index variations $\ge 0.45$ $\%$ in the aqueous medium, shifting the plasmon by $1$ nm.
In the low concentration regime, the reduction for the static dielectric constant ($\epsilon'$) reads:\cite{adar18}
\be
\epsilon' \; = \; \epsilon_w - \; \gamma c_{_\pm} \; + \; \zeta c^{\f{3}{2}}_{_\pm}
\label{eqn:d19}
\ee
where $w$ refers to water (solvent), $c_{\pm}$ is the electrolyte concentration, $\gamma$ and $\zeta$ are linear and nonlinear decrement terms.
With a surfactant concentration $c_{_\Sigma} = 15$ mM (CTAC), one may limit to the lowest-order term:
\be
\gamma \; = \; 2 a^3 \epsilon_w \left( 1 + \f{4 l_{_B}}{3 a} \right)
\label{eqn:d20}
\ee
$a$ being equivalent to the volume fraction of dissociated species in solution (i.e. $\phi_{\pm} = 2 c_{\pm} a^3$) and $l_{_B}$ denoting Bjerrum's length.\\
Eq. (\ref{eqn:d19}) is applicable to the initial dielectric constant of the bulk medium, composed by surfactant (subindex $\Sigma$, $\epsilon_{_\Sigma} = 2.074$) and solvent:
\be
\epsilon_m \; = \; \phi_w \epsilon_w + \; \phi_{_\Sigma} \epsilon_{_\Sigma}
\label{eqn:d21}
\ee
The surfactant volume fraction in solution $\phi_{_\Sigma} \approx (1 - \alpha_{_\Sigma}) c_{_\Sigma} M_{_\Sigma} / \rho_{_\Sigma}$ may be estimated from its dissociation degree\cite{sep85} ($\alpha_{_\Sigma} \approx 0.4$), molecular weight ($M_{_\Sigma} \approx 320$ g/mol), mass density ($\rho_{_\Sigma} \le 0.97$ g/cm$^3$), to get $\phi_{_\Sigma} \le 3.2 \cdot 10^{-3}$.
Now, let $a^3 = \f{1}{2} (a^3_{-} + a^3_{+})$, one obtains from equation (\ref{eqn:d19}):
\be
\f{\epsilon'_m}{\epsilon_m} \; \approx \; 1 - \left( 1 + \f{4 l_{_B}}{3 a} \right) \phi_{\pm}
\label{eqn:d22}
\ee
in which, as ionic species are chloride Cl$^-$ and hexadecyltrimethylammonium CTA$^{+}$, an estimate $a_{+}/a_{-} \sim 2 \div 3$ should apply ($a_- \approx 3.6$ {\AA},\cite{adar18} the length of CTA$^{+}$ $>3$ nm\cite{chao12}).
This implies $\phi_{\pm} \sim (0.9 \div 2.5) \cdot 10^{-3}$, $4 l_{_B}/(3 a) \sim 1.07 \div 1.58$ and, finally $\epsilon'_m \approx (0.995 \div 0.997) \; \epsilon_m$, which corresponds to a corrected $n' \approx 1.327 \div 1.329$.\\
Therefore, as the refractive index reduction is $< 0.25$ $\%$ in absolute value, electrolyte effects can be ruled out.
With the same physical chemistry parameters, the dielectric reduction of the surfactant shell required to get a refractive index lowering of $0.45$ $\%$ amounts to $\epsilon_{_\Sigma} = (0.94-2.05)$, i.e. to an average reduction of $28$ $\%$ ($\epsilon_{_\Sigma} \approx 1.49$).
This is also highly unlikely.

\subsection*{\rm \bf E. Geometric Polydispersity}

\renewcommand{\theequation}{E.\arabic{equation}}
\setcounter{equation}{0}

Plasmon wavelengths may also vary as a consequence of geometric fluctuations of particle radius and shape descriptor (or circularity).
Though a fraction of irregular units could surely shift the peak position, for a geometrically symmetric distribution this turns out to produce a contained red-shift.\\
We have built a bimodal nanoparticle distribution, its joint probability for aleatory radius and size being factorized into independent Gaussian functions:
\be
p_{ij} (R = r, L = l) \; = \; p_i (R = r) \; p_j (L = l)
\label{eqn:e23}
\ee
Using the former statistical relationship between radius and shape descriptor, the electromagnetic response was simulated for a statistical ensemble of nine particles of the sample with the largest $\langle L \rangle$.
To compute each particle contribution to the average extinction spectra, the two distributions $p_i$ and $p_j$ were discretized into a three-bar histogram, each bar being centered at $\langle X \rangle - \frac{3}{2} \sigma$, $\langle X \rangle$ and $\langle X \rangle + \frac{3}{2} \sigma$.
Particle statistical weights then were calculated by the normal distribution value in the middle of the bar, except from the one centered at $\langle X \rangle$.
In this case, the values of the Gaussian at $X = \pm \frac{1}{2} \sigma$ were set to avoid overestimation of the $\langle X \rangle$ contribution.\\
Figure (\ref{fig:poly-a}) reports the individual extinction cross-sections, from which the weighted average yields the polydisperse spectra of Figure (\ref{fig:poly-b}), superimposed for convenience to the single-particle profile.
While the full width at half maximum is meaningfully increasing with the geometric dispersion, from $67 \pm 1$ to $77 \pm 1$ nm, the moderate wavelength change we get is positive, $\Delta \lambda < 1.5$ nm, and thus is unsuitable to explain our discrepancy, especially for the largest particles.
An appreciable blue-shift would rather require an asymmetric distribution of more circular particles.
Such an hypothesis, though being not fully unlikely, cannot however be stated offhand and, presently, we haven't arguments for it.

\begin{figure}[h]
	\centering
	\includegraphics[scale=0.42]{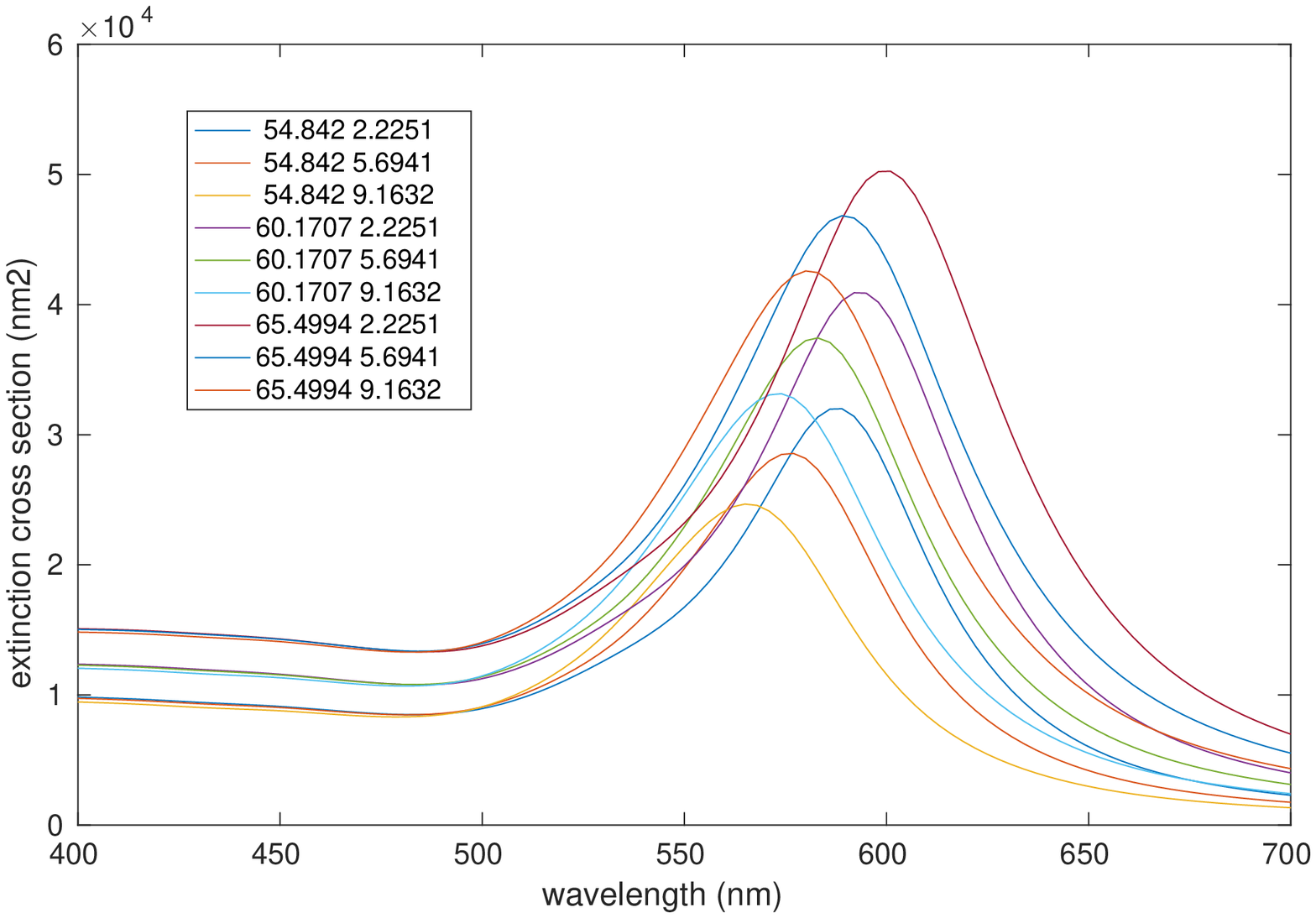}
	\caption{Individual extinction cross-sections for the polydisperse particle ensemble.}
	\label{fig:poly-a}
\end{figure}

\begin{figure}[h]
	\centering
	\includegraphics[scale=0.4]{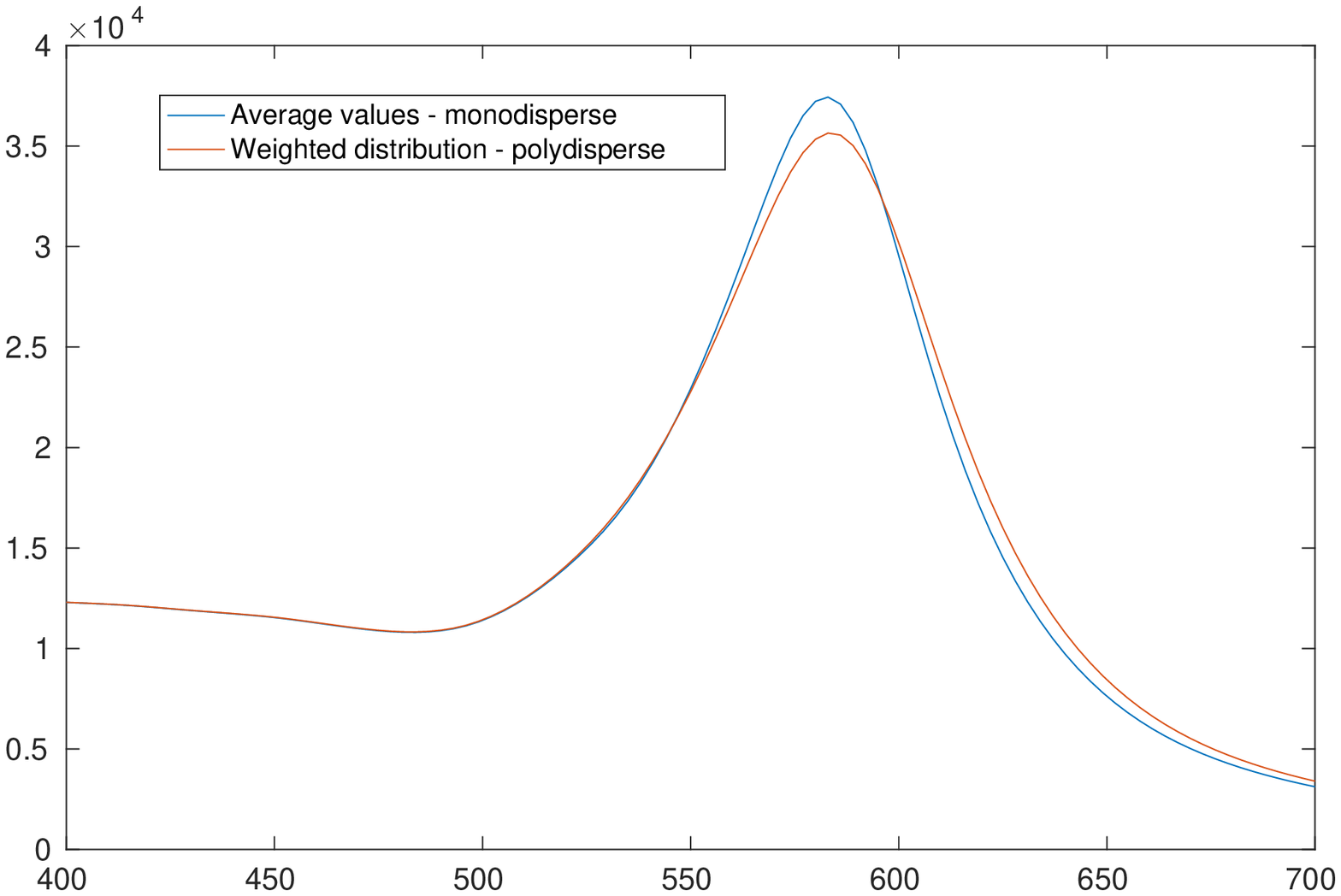}
	\caption{Average polydisperse versus monodisperse cross-sections.}
	\label{fig:poly-b}
\end{figure}

\subsection*{\rm \bf F. Free and Bound Surface Charges}

\renewcommand{\theequation}{F.\arabic{equation}}
\setcounter{equation}{0}

Chemically reducing a salt solution into metal colloids generally retains some surface charge that, for Au, is normally negative.
Correspondingly, a diffuse ion cloud surrounding the particle takes place as a charge compensation, building up an electrochemical/metal capacitance.
If ionic surfactants then are employed as colloid stabilizers, an excess charge may be bound at or in proximity to the surface by means of ion adsorption.
Here, a CTAC bilayer long $3.5$ nm is formed, CTA$^+$ being adsorbed at Au$-$O$^-$ equilibrium sites:\cite{muto07}
\be
{\rm Au-OH} \; \rightleftharpoons \; {\rm Au-O}^- + \; {\rm H}^+
\label{eqn:f25}
\ee
and yielding a $\zeta$ potential $\sim 50$ meV in our concentration regimes (CMC $\sim 1$ mM).\\
To modify the dielectric function of Au accordingly, let's start from the bulk frequency-dependent function, $\epsilon^b_{\rm Au} \left( \omega , \omega_b\right)$.
In the optical range, it writes:
\begin{equation}
\epsilon_{\rm Au}^b \left( \omega \right) \; = \; \epsilon_\infty + \epsilon_{\rm int} \left( \omega \right) - \frac{\omega_{b}^2}{\omega \left(\omega + {\rm i} \gamma \right)} 
\end{equation}
where $\epsilon_\infty$ is a constant contribution from polarization mechanisms resonating at photon energies above the spectral range of interest, $\epsilon_{\rm int}$ accounts for interband transitions and the last term is Drude's, expressed by the plasma frequency ($\omega_{\rm p} \equiv \omega_b$) and damping constant ($\gamma$).
Best fitting the ligature optical constants for Au single crystals \cite{olmon2012optical} leads to $\omega_{\rm p} = 8.32$ eV and $\gamma = 45$ meV.
Olmon et al's data were preferred to Johnson \& Christy's for a better reproduction of extinction widths (see the example in Figure \ref{fig:jc}).

\begin{figure}[h]
	\centering
	\includegraphics[scale=0.27]{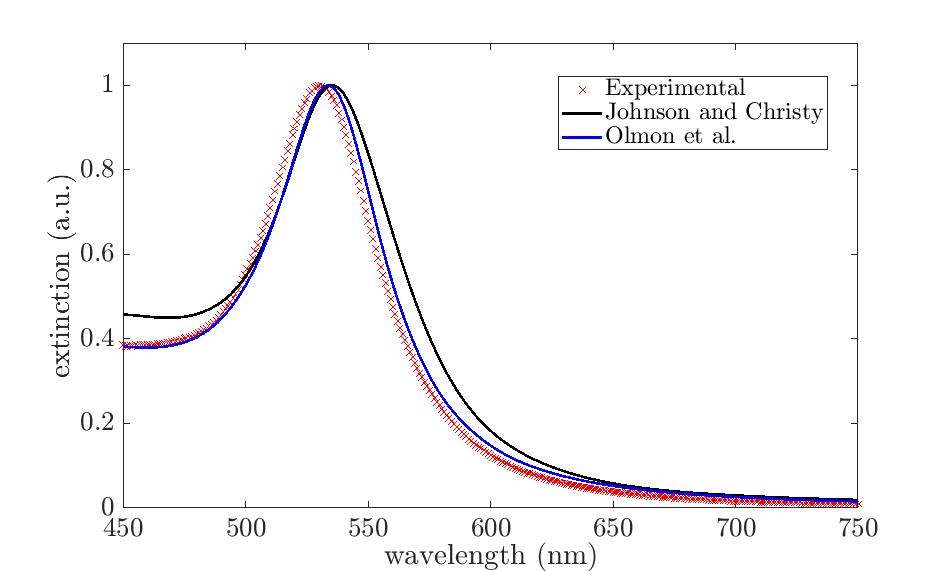}
	\caption{Prediction with Olmon et al's data versus Johnson \& Christy's for the sample with ${\rm S} = 0.58$.}
	\label{fig:jc}
\end{figure}

\noindent An excess free charge yields an additional contribution to the surface plasma frequency ($\omega_s$) linked to the surface charge, which leads to the following expression for the corrected $\epsilon$:
\begin{equation}
\epsilon_{\rm Au}^c \left( \omega \right) = \epsilon_\infty + \epsilon_{\rm int} \left( \omega \right) - \frac{\omega_{s}^2 + \omega_{b}^2}{\omega \left( \omega + {\rm i} \gamma \right)} 
\end{equation}
As the LSP resonance occurs at the frequency where the real part of the dielectric function equals a (real-valued) particle eigenvalue (see Theoretical section), $\omega_s$ can be derived from:
\be
\Re [ \epsilon_{\rm Au}^b \left (\omega_{\rm num} \right) ] \; = \; \Re [ \epsilon_{\rm Au}^c ( \omega_{\rm exp} ) ]
\ee
$\omega_{\rm exp}$ being the experimentally determined frequency of LSP resonance and $\omega_{\rm num}$ the one numerically computed from the bulk dielectric function.
Therefore:
\begin{equation}
\omega_s^2 \; = \; \omega_b^2 \left( \frac{\omega_{\rm exp}^2 - \omega_{\rm num}^2}{\omega_{\rm num}^2 + \gamma^2} \right) - \Re [ ( \epsilon_{\rm int} \left( \omega_{\rm num} \right) - \epsilon_{\rm int} ( \omega_{\rm exp} )] ( \omega_{\rm exp}^2 + \gamma^2 )
\label{eqn:f27}
\end{equation}
On the other hand, the ratio $\omega_{s}^2/\omega_{b}^2$ may be developed by the numbers of free charges in the metal bulk ($N_b$), at the surface ($N_s$) and of effective excess charges at the interface ($N_{\iota}$), $\omega^2_b \propto N_b$ and $\omega^2_s \propto | N_s - N_{\iota} |$. If $\rho_{s / b}$ and $\rho_{\iota / b}$ denote two effective surface densities for $N_s$ and $N_\iota$, both normalized to the reference electron density of Au ($59$ nm$^{-3}$),\cite{ash1976} one obtains:
\be
\f{\omega_{s}^2}{\omega_{b}^2} \; \approx \; - \; \rho_{\iota / b} \; s \; + \; \rho_{s / b} 
\label{eqn:f28}
\ee
the particle specific surface ($s$) accounting for the ratio between surface and volume.
In particular, it turns out $s = - 0.108 \; {\rm S} + 0.178$ and the radii of spherical particles with equivalent volume behaves as $R^* = (29.92 \; {\rm S} + 11.80)$ nm ($R^2 = 0.70$).
Figure (\ref{fig:excesswp}) compares the experimental with expected numerical values, confirming the above linear trend with best fit parameters $\rho_{s / b} = 0.32$ and $\rho_{\iota / b} = 1.73$ nm ($R^2 = 0.78$, blue line).
Observe that $N_s / N_b < 1$ and $N_{\iota}/N_b = (16 \div 28) \cdot 10^{-2}$, which is a reasonably small fraction of effective excess charges ($N_\iota < N_s$).\\
If Drude's is the dominant contribution, a straightforward formula for the excess plasma frequency is:
\be
\f{\omega_{s}^2}{\omega_{b}^2} \; = \; \f{\omega_{\rm exp}^2 - \; \omega_{\rm num}^2}{\omega_{\rm num}^2}
\label{eqn:f28b}
\ee
This approximation was exploited to get the plasmon shift upon excess charges in Drude-like spherical particles,\cite{rost96, bohr77} and to quantify them in Au nanorods,\cite{mulvaney2006drastic} where Drude's term dominates the dielectric behavior and LSP resonances lie in the near-infrared range.
Here, however, they occur at photon energies where interband transitions aren't negligible.
In fact, while equation (\ref{eqn:f27}) refines the simulations to a very good agreement with experiments, equation (\ref{eqn:f28b}) alone would return $\rho_{s / b} = 0.24$ and $\rho_{\iota / b} = 1.36$ nm ($R^2 = 0.74$), amending the plasmon position by $(3-6)$ nm lesser than in former predictions.
Finally, one can get the overall plasma frequency's increase due to charge injection, $\omega^2_p/ \omega^2_b = (1 + \xi) \; \omega^2_b$, where $\xi$ is the ratio in equation (\ref{eqn:f28b}).
The numerical computations were best fitted by $\xi = 0.213 \; {\rm S} - 0.041$ ($R^2 = 0.88$), which were used with $R^* = R^* ({\rm S})$ in retarded calculations (equation \ref{eqn:last}).\\
To sum up, the anomalous red-shift correcting our data is mostly ascribed to equation (\ref{eqn:f27}).
Two main approximations were made, a size-independent density $\rho_{s / b}$ and an interface area roughly estimated by the particle surface.
To get a better accuracy, the linear behavior was corrected into a heuristic power law of the particle specific surface.
A second best fit thus was performed in Figure (\ref{fig:excesswp}) upon replacing the right side of equation (\ref{eqn:f28}) with $\beta_{_1} s^{\beta} + \beta_{_2}$, getting $|\; \beta_{_1}| = 2.7 \cdot 10^{-4}$, $\beta = -2.68$, $|\; \beta_{_2}| = 0.02025$ ($R^2 = 0.83$, green line).
It turns out the LSP wavelength of the largest cubic particles (${\rm S} = 0.81$) is refined to $\lambdabar \; {\rm (th)} = 554$ nm, reducing the highest relative uncertainty in our distribution from $2.2 \%$ to $1.6 \%$.
This, however, does not change anything in the physical and chemical picture behind equations (\ref{eqn:f27}-\ref{eqn:f28}).

\begin{figure}[h]
	\centering
	\includegraphics[scale = 0.32,left]{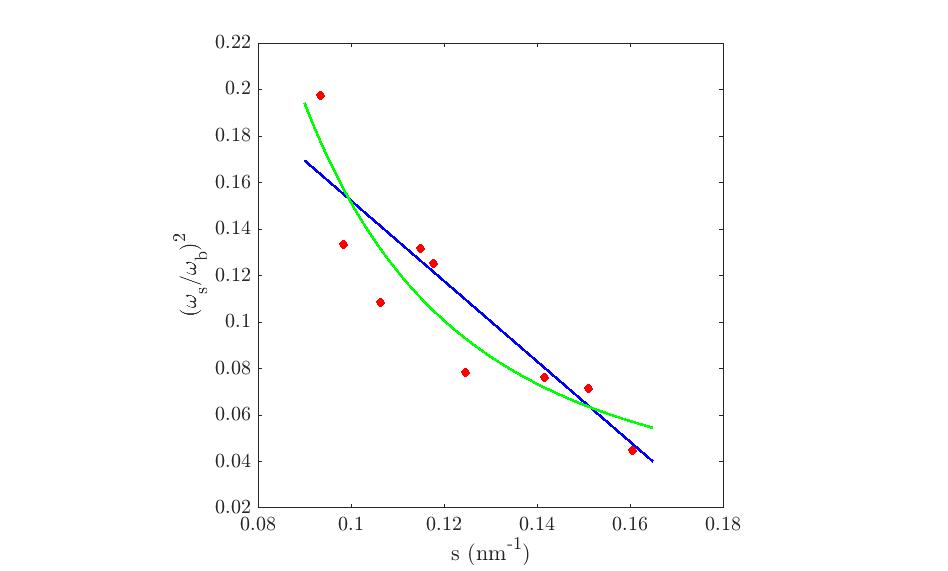}
	\caption{Excess Drude's plasma frequency correction versus particle specific surface.
	Blue line reports the best fit $(\omega_{s}/\omega_{b})^2 \; \approx \; - \; \rho_{\iota / b} \; s \; + \; \rho_{s / b}$.
	In green is the power-law correction illustrated in the text.}
	\label{fig:excesswp}
\end{figure}	

\subsection*{G. Summary of all LSP resonance data}

\saltino

\renewcommand{\theequation}{G.\arabic{equation}}
\setcounter{equation}{0}

All plasmon resonance predictions are grouped in Table (\ref{tbl:syn}), i.e. experimental (ex), the values coming from uncorrected and geometrically corrected shape descriptors (unc and sc), and the ones undergoing geometric and charge corrections without (scc) and with (th) interband transitions.
Note that, while the fifth column (qs) reports the quasistatic (geometrically corrected) values, any other wavelength $\lambdabar$ underwent retardation.

\begin{table*}[h]
  \caption{Summary of LSP resonance predictions (wavelengths $\lambdabar$ in nm).}
  \label{tbl:syn}
\begin{tabular*}{\textwidth}{@{\extracolsep{\fill}}lllllll}

    \hline
		 {\rm S} & $\lambdabar$ \; ({\rm ex}) & $\lambdabar$ \; ({\rm unc}) & $\lambdabar$ \; ({\rm sc}) & $\lambdabar$ \; ({\rm qs}) & $\lambdabar$ \; ({\rm scc}) & $\lambdabar$ \; ({\rm th}) \\

		\hline
0.51 & 530 & 558 & 544 & 536 & 535 & 531 \\
0.41 & 525 & 552 & 538 & 530 & 531 & 528 \\
0.26 & 523 & 548 & 531 & 526 & 527 & 524 \\
0.67 & 537 & 575 & 562 & 545 & 546 & 540 \\
0.58 & 530 & 565 & 553 & 540 & 539 & 534 \\
0.38 & 526 & 555 & 540 & 530 & 528 & 524 \\
0.81 & 545 & 597 & 585 & 556 & 563 & 557 \\
0.55 & 533 & 570 & 558 & 539 & 540 & 534 \\
0.44 & 528 & 561 & 548 & 533 & 532 & 527 \\
    \hline
  \end{tabular*}
\end{table*}

\section*{\rm \bf 6. Mathematical Details}

\subsection*{\rm \bf H. Basic References to Crystal (Ligand) Field Theory (CFT)\cite{bet29,figgis00,scl69,per81}}

\renewcommand{\theequation}{H.\arabic{equation}}
\setcounter{equation}{0}

CFT describes how electrons of a central ion (normally, a transition metal atom) interact with a group of ligands, represented as pointwise negative charges.
Results will depend on the spatial symmetries of the orbital involved, settled by a linear superposition of eigenfunctions $\psi_{k} = \psi_k ({\bf r})$ ($k = 1, \; ... \; m$) of the unperturbed Hamiltonian, and of the perturbation potential, generated by the crystal field ${\rm U} = {\rm U} ({\bf r})$.
Key quantities thus are the matrix elements ${\rm U}_{ks} = (\psi_k, {\rm U} \psi_s)$ of the secular determinant:
\be
\parallel {\rm U}_{k s} - \varepsilon \delta_{k s} \parallel \; = 0
\label{eqn:h31}
\ee
where, for a $m$-fold degenerate level, $\varepsilon_k$ ($k = 1, \; ... \; m$) are eigenvalues of the energy perturbation.
For weak crystal fields, spin-orbit coupling and Coulomb's interactions with all ligands may be neglected.
We are therefore left with a crystal field potential ${\rm U}$ expressed as a sum of electron-electron operators for a given crystal symmetry and orbital term, with the single-electron wavefunction provided as usual by $|n \ell m \rangle = R_{n \ell} Y_{\ell m}$.\\
Upon cubic symmetries, $S$ states undergo no change and $P$'s don't split, as it may be argued from the form of matrix elements in such cases.
$D$ terms then split into two, $F$'s into three, $G$ and $H$'s into four levels each.
In the specific case e.g. of $D$ states in an octahedral field, one obtains: 
\be
{\rm U} (O_h, d) \; = \; 12 \sqrt{\pi} \; \f{q_i}{l} \left\{ Y_{0 0} \; + \; \left( \tfrac{7}{36} \right) \f{\overline{r^4}}{l^4} \left[ Y_{4 0} + \sqrt{\tfrac{5}{14}} \left( Y_{4 4} + Y_{4 -4} \right) \right] \right\}
\label{eqn:h32}
\ee
$q_i$ denoting the ion charge, $l$ the ligand distance from the origin and, here:
\be
\overline{r^q} \; = \; \int^\infty_{0} r^{q + 2} |R_{n 2}|^2 dr
\label{eqn:h33}
\ee
estimating the mean $q$th-power radius of a $d$ electron.
In equation (\ref{eqn:h32}), $Y_{0 0}$ yields an equal energy change for all orbitals (i.e. a $\overline{r^0}$ term).
It translates the bringing up to the ion of a homogeneous spherical shell of (negative) charge.
The relative energies stem in diagonal setting from letting the basis to transform according to the irreducible representations $E_g$ and $T_{2g}$ of $O_h$, getting the so-called cubic harmonics for $\ell = 2$ (principal number $n$ is omitted):
\be
\psi_{_1} (E_g) \; = \; | 2 0 \rangle
\;,\;\;
\psi_{_2} (E_g) \; = \; \tfrac{1}{\sqrt{2}} \; (|2 2 \rangle + |2 - 2 \rangle)
\label{eqn:h34}
\ee
\be
\psi_{_3} (T_{2 g}) \; = \; - \; |2 1 \rangle
\;,\;\;
\psi_{_4} (T_{2 g}) \; = \; |2 -1 \rangle
\;,\;\;
\psi_{_5} (T_{2 g}) = \; \tfrac{1}{\sqrt{2}} \; (|2 2 \rangle - |2 - 2 \rangle)
\label{eqn:h35}
\ee
so that $\varepsilon (E_g) \equiv {\rm U}_{1 1} = {\rm U}_{2 2}$ and $\varepsilon (T_{2 g}) \equiv {\rm U}_{3 3} = {\rm U}_{4 4} = {\rm U}_{5 5}$ turn out to be the related two-fold and three-fold splitting.
Such calculations, involving Clebsh-Gordan coefficients for three spherical harmonics ($3 \ell$-symbols), can be summarized as $2 \varepsilon (E_g) = - 3 \varepsilon (T_{2 g}) > 0$.
For tetrahedral fields this phenomenology is reverted, as one finds ${\rm U} (T_d, d) = - (4/9) {\rm U} (O_h, d)$, but the main concepts are unchanged.\\
When $f$ orbitals are regarded, it is necessary to expand the electrostatic potential up to $2 \ell = 6$, i.e.:
\be
{\rm U} (O_h, f) \; = \; {\rm U} (O_h, d) \; + \;
\sqrt{\tfrac{\pi}{13}} \left( \tfrac{3}{2} \right) q_i \; \f{\overline{r^6}}{l^7}
\left[ Y_{6 0} + \sqrt{\tfrac{7}{2}} \left( Y_{6 4} + Y_{6 -4} \right) \right]
\label{eqn:h36}
\ee
and recalculate the secular equation in the irreducible representations of the cubic group.
This approach is promptly generalized to any spectroscopic term ($G$, $H$, etc.) and, by virtue of equations (\ref{eqn:h32}, \ref{eqn:h36}), lies the basis of equation (\ref{eqn:10}) of the text.

\vspace{1mm}

\subsection*{\rm \bf I. Mean Value Theorem for Definite Surface Integrals}

\vspace{1mm}

\renewcommand{\theequation}{I.\arabic{equation}}
\setcounter{equation}{0}

Consider a function $h: A \rightarrow \Re$, where $h \in C (A)$ and $A \in \Re^2$ is a compact and rectifiable set.
The one-dimensional mean value theorem generalizes in two dimensions as:
\be
\iint_{A} h ds \; = \; h ({\bf a}) \iint_{A} ds
\label{eqn:i29}
\ee
where the integral on the right is the area of $A$ and point ${\bf a} \in A$ is, generally, not unique.
An equivalent expression clearly holds for any continuous vector function ${\bf h}$ defined onto an arbitrary disc $A_{d \;}$:
\be
\iint_{A_d} ({\bf n} \cdot {\bf h}) ds \; = \; {\bf n} \cdot {\bf h} ({\bf a}_{d}) \iint_{A_d} ds
\label{eqn:i30}
\ee
still with ${\bf a}_{d} \in A_{d}$ and ${\bf n}$ being the normal unit vector that points outward the surface.

\subsection*{\rm \bf L. Best Fitting Eigenvalues vs Shape Descriptor}

\renewcommand{\theequation}{L.\arabic{equation}}
\setcounter{equation}{0}

Equation (\ref{eqn:13}) in the text is a fourth-degree polynomial in ${\rm S}$, with coefficients:
\be
\begin{array}{lll}
a_{_0} \; = \; - \; \mu_{\ell m} z^2_{\underline{k} s}\\
a_{_1}  \; = \;  (2 \mu_{\ell m} - \; \sigma_{\ell m}) z^2_{\underline{k} s} - 2 z_{\underline{k} b} z_{\underline{k} s} \mu_{\ell m}\\
a_{_2} \; = \; - \; \chi_{\ell m} z^2_{\underline{k} s} + \mu_{\ell m} (4 z_{\underline{k} b} z_{\underline{k} s} - z^2_{\underline{k} b} - z^2_{\underline{k} s}) + \sigma_{\ell m} (z_{\underline{k} s} - 2 z_{\underline{k} b}) z_{\underline{k} s}\\
a_{_3} \; = \; (\sigma_{\ell m} - \chi_{\ell m} - \mu_{\ell m}) 2 z_{\underline{k} b} z_{\underline{k} s} + (2 \mu_{\ell m} - \sigma_{\ell m}) z^2_{\underline{k} b}\\
a_{_4} \; = \; (\sigma_{\ell m} - \chi_{\ell m} - \mu_{\ell m}) z^2_{\underline{k} b}
\end{array}
\label{eqn:l37}
\ee
Now, for any quintuple $\{ a_{i} \}$, rescale $a'_{i} = a_{i}/z^2_{\underline{k} s}$.
The quantities $\mu_{\ell m}$, $2 \mu_{\ell m} - \; \sigma_{\ell m}$, $\sigma_{\ell m} - \chi_{\ell m} - \mu_{\ell m}$, $z_{\underline{k} b}/z_{\underline{k} s}$ then follow from three of the above equations as functions of $a'_{i}$, the fourth completing the solution by means of $z_{\underline{k} s}$.
This proves equation (\ref{eqn:13}) to belong to $P_{_4} [0,1]$.\\
We best fitted all eigenvalues in Figure (\ref{fig:eigenmodes}), with coefficients ranging in $\sim \pm 10^{5}$, and the agreement was excellent (Tables \ref{tbl:fit-a} and \ref{tbl:fit-b}).
To reduce the extent of possible numerical uncertainties due to wedge- or cusp-like effects (${\rm S} \rightarrow 1^-$), all interpolations were conducted twice, the first time unconstrained ($u$) and the second restricted to ${\rm S} \le 0.9$ ($r$).
\begin{table}[h]
  \caption{Details of best fits for semicube's eigenvalues ($\lambda_{\underline{k} b}$).}
  \label{tbl:fit-a}
  \begin{tabular*}{0.48\textwidth}{@{\extracolsep{\fill}}llllllll}
    \hline
        $\ell$ & {\rm deg} $(m)$ & $\chi_{\ell m}$ & $\sigma_{\ell m}$ & $\mu_{\ell m}$ & $z_{\underline{k} s}$ & $z_{\underline{k} b}$ & $R^2$ \\
		\hline
1 & 3 & 1.34 & 0.67 & 1.42 & 1.73 & 0.01 & 0.997 \\
2 & 2 & 0.95 & 0.71	& 0.84	& 1.44 & 0.07 &	0.999 \\
  & 3 & 0.63 &	0.92 &	1.01 &	1.26& 1.28 & 0.999 \\
3 & 1 & 0.62 & 0.98	& 0.98	& 1.15	& 1.39	& 0.999 \\
  & 3 & 0.70 & 0.88	& 0.66 & 1.22 & 0.82 & 0.997 \\
  & 3 & 0.64 & 0.85	& 0.82 & 1.26 &	0.60 & 0.998 \\
4 & 3 & 0.56 & 0,99	& 0.49 & 1.12 & 1.11 & 0.998 \\
  & 2 & 0.60 &	1.28 & 0.76 & 0.96 & 0.87 &	0.979 \\
  & 3 & 0.49 &	1.13 & 0.58	& 1.04 & 1.05 & 0.999 \\
  & 1 & 0.70 & 0.86	& 0,63 & 1.17 & 1.06 & 0.997 \\
5 & 3 & 0.60 & 1.27	& 0.61 & 0.96 & 1.02 & 0.996 \\
  & 2 & 0.53 & 1.30	& 0.62 & 0.96 & 1.06 & 0.998 \\
  & 1 & 0.41 & 0.94	& 0.52 & 1.10 &	1.03 &	0.975 \\
  & 2 & 0.48 & 1.21	& 0.61 & 0.97 &	1.05 &	0.975 \\
  & 1 & 0.57 & 0.98	& 0.65 & 1.10 &	0.84 &	0.997 \\
  & 2 & 0.48 & 0.82	& 0.54 & 1.20 & 0.92 &	0.997 \\
    \hline
  \end{tabular*}
\end{table}

\begin{table}[h]
  \caption{Details of restricted best fits for $\lambda_{\underline{k} b}$ ($^{a \;}$the value taken by $z_{\underline{k} b}$ for $\ell$ $= 1$ was $-3.2 \cdot 10^{-3}$).}
  \label{tbl:fit-b}
  \begin{tabular*}{0.48\textwidth}{@{\extracolsep{\fill}}llllllll}
    \hline
        $\ell$ & {\rm deg} $(m)$ & $\chi_{\ell m}$ & $\sigma_{\ell m}$ & $\mu_{\ell m}$ & $z_{\underline{k} s}$ & $z_{\underline{k} b}$ & $R^2$ \\
		\hline
1 & 3 & 1.27 & 0.63 &	1.42 &	1.77 &	$\approx$ 0$\textsuperscript{\emph{a}}$ & 0.999\\
2 & 2 & 0.82 &	0.53	& 0.67 &	1.68 &	0.07 &	0.999\\
 & 3 & 0.78 &	0.83	& 1.29	& 1.34	& 0.89 & 0.999\\
3 & 1 & 0.58 & 0.74	& 0.97	& 1.34	& 1.26	& $\approx$ 1\\
 & 3 & 0.71	& 0.60	& 0.70	& 1.49	& 0.46 &	0.999\\
 & 3 & 0.59	& 0.80 &	0.75	& 1.30 &	0.64	& 0.997\\
4 & 3 & 0.61&	0.97 &	0.62	& 1.14 &	0.95	& 0.998\\
  & 2 & 0.45 &	0.80 &	0.67 &	1.22 & 0.84	& 0.960\\
  & 3 & 0.48	& 1.12 & 0.56	& 1.05 &	1.07	& 0.997\\
  & 1 & 0.74 &	0.54	& 0.69	& 1.50 & 0.61	& 0.999\\
5 & 3 & 0.47	&1.01	 & 0.46	& 1.07	& 1.17	& 0.993\\
 & 2 & 0.40	&1.00	 & 0.45	& 1.09	& 1.25 &	0.997\\
 & 1 & 0.44	& 0.90 &	0.65	& 1.13 &	0.95 &	0.973\\
 & 2 & 0.43	& 0.94 &	0.70	& 1.11 &	0.93	& 0.954\\
 & 1 & 0.65	& 1.06 &	0.77	& 1.06& 0.74	& 0.995\\
 & 2 & 0.58	& 0.95 &	0.69	& 1.12 &	0.79	& 0.995\\
		\hline
    \hline
  \end{tabular*}
\end{table}

\begin{figure}[!h]
	\centering
	\includegraphics[scale=0.2]{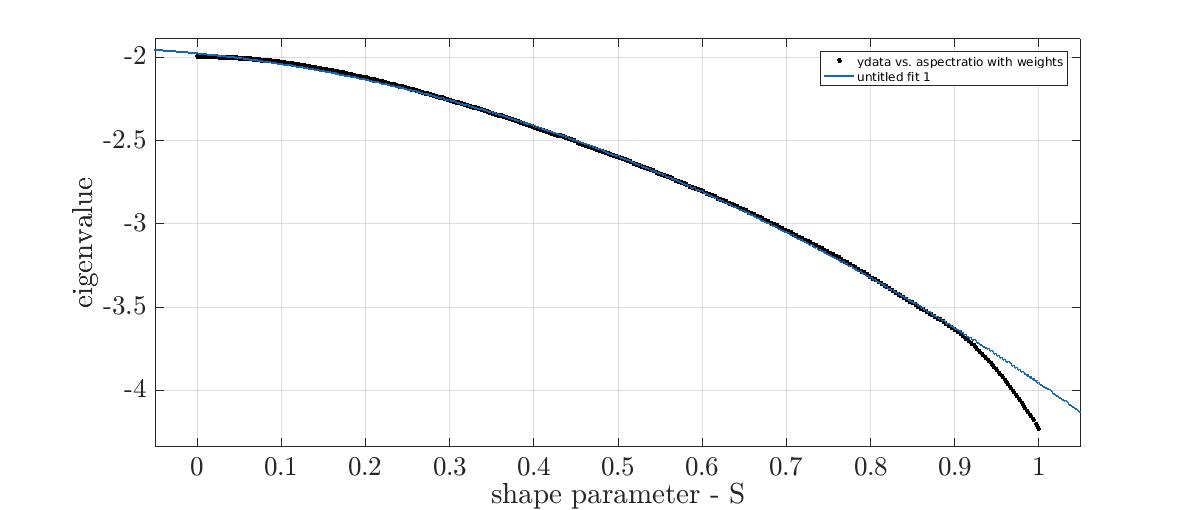}
	\caption{Restricted polynomial for $\ell = 1$.}
	\label{fig:a}
\end{figure}
\begin{figure}[!h]
	\centering
	\includegraphics[scale=0.2]{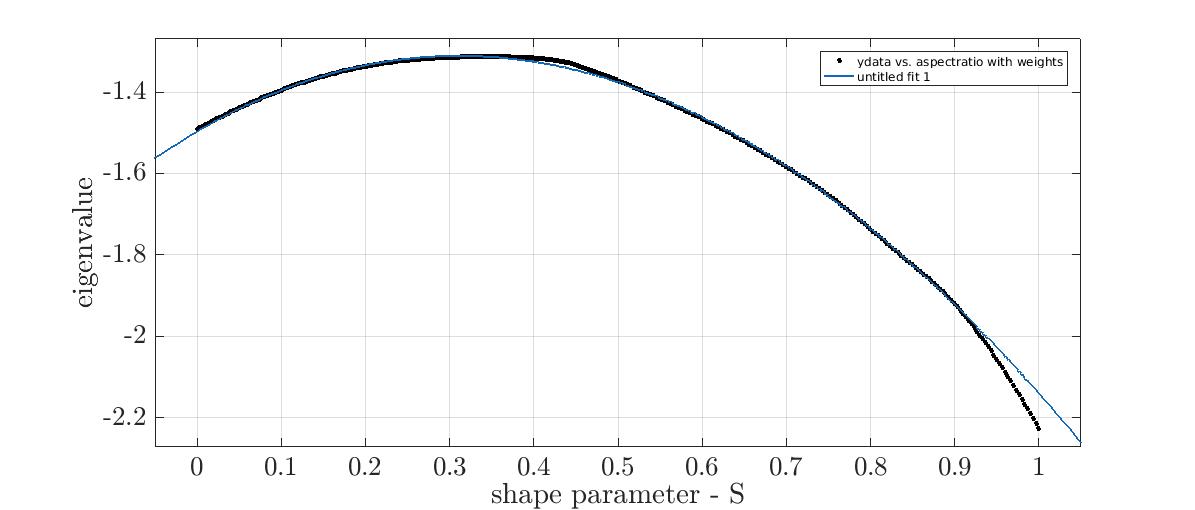}
	\caption{Restricted polynomial for $\ell = 2$ ($E_g$).}
	\label{fig:b}
\end{figure}
\begin{figure}[!h]
	\centering
	\includegraphics[scale=0.2]{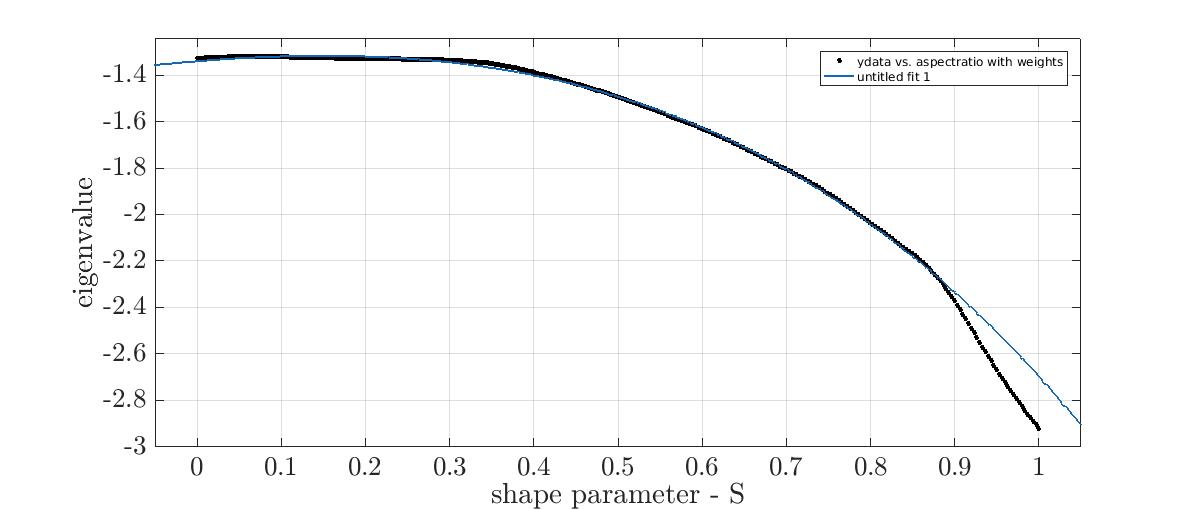}
	\caption{Restricted polynomial for $\ell = 3$ ($T_{1g}$).}
	\label{fig:c}
\end{figure}
\begin{figure}[!h]
	\centering
	\includegraphics[scale=0.2]{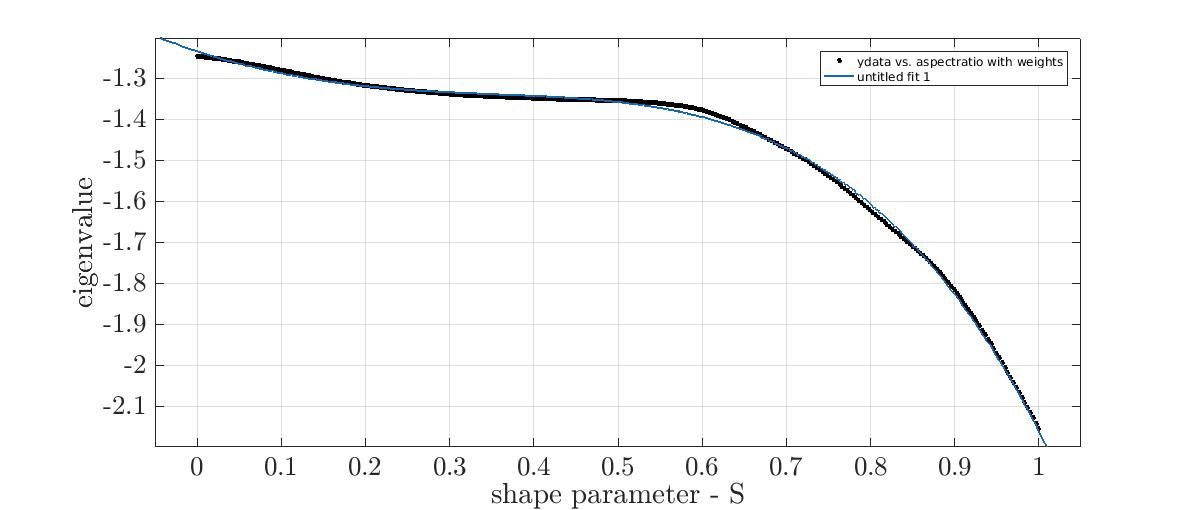}
	\caption{Restricted polynomial for $\ell = 4$ ($T_{1g}$).}
	\label{fig:d}
\end{figure}

\newpage

\noindent In this way, cube eigenvalues could be extrapolated for ${\rm S} \rightarrow 1^-$ by weighting over the coefficients of determination ($R^2 > 0$) derived in both cases to:
\be
\lambda_{\underline{k} c} \; = \; \f{R^2_{\underline{k} u} \; \lambda_{\underline{k} c,u} + R^2_{\underline{k} r} \; \lambda_{\underline{k} c,r}}{R^2_{\underline{k} u} + R^2_{\underline{k} r}}
\label{eqn:i38}
\ee
obviously with ($j = r, u$):
\be
\lambda_{\underline{k} c, j} = - \chi_{_{\ell m, j}} (z_{\underline{k} m, j} + z_{\underline{k} s, j})^2
\label{eqn:i39}
\ee
These results are shown in Table (\ref{tbl:tab2}) of the main text.
Examples of restricted fits instead are in Figures (\ref{fig:a} - \ref{fig:d}).
Unrestricted plots, however, didn't show significant differences, as it is clear from the values taken by $R^2$ in the last two tables.

\subsection*{\rm \bf M. Retardation Calculations}

\renewcommand{\theequation}{M.\arabic{equation}}
\setcounter{equation}{0}

i.) Equation (\ref{eqn:n}) comes from expanding the exponential in equation (\ref{eqn:rho}) and separating the contributions to the vector potential (${\rm R} \equiv |{\bf r} - {\bf r}'|$):
\be
{\bf A}^{\rm \; ret}_{\bf \rho} \; = \; \sum_{n \; \ge \; 0} \int \left( \f{({\rm i} k)^{n - 1}}{n !} {\rm R}^{n - 2} \; - \; \f{({\rm i} k)^n}{n !} {\rm R}^{n - 1} \right) \rho ({\bf r}') \; \widehat{\bf R} \; d^3 r'
\ee
giving:
\be
= \; - \; \f{{\rm i} }{k} \int \f{\rho ({\bf r}')}{{\rm R}^{2}} \; \widehat{\bf R} \; d^3 r' \; + \; \sum_{n \; \ge \; 1} ({\rm i} k)^{n - 1} \left( \f{1}{n!} - \f{1}{(n - 1) !} \right) \int \f{\rho ({\bf r}')}{{\rm R}^{2 - n}} \; \widehat{\bf R} \; d^3 r'
\ee
The second integral identifies ${\bf I}_n$ in equation (\ref{eqn:n}), while the first:
\be
\int \f{\rho ({\bf r}')}{{\rm R}^{2}} \; \widehat{\bf R} \; d^3 r' \; = \; - \; \nabla \int \f{\rho ({\bf r}')}{{\rm R}} \; d^3 r' \; = \; - \; \nabla V
\ee
returns the instantaneous electric field component in Coulomb's gauge.\\
\\
ii.) To get the numerator in equation (\ref{eqn:13-b}), the sum:
\be
\f{4 \pi}{R} \sum^{\infty}_{\ell = 0} \sum^{\ell}_{m = - \ell}
\f{1}{2 \ell + 1} \iint Y^*_{\ell m} (\theta', \varphi') \; \widehat{p}_{\bf n} Y_{\ell m} (\theta, \varphi) d s d s' \; =
\ee
needs to be worked out for a spherical particle of radius $R$, with $\widehat{p}_{\bf n}$ only including the geometrical part of the dipole moment, $\widehat{p}_{\bf n} = L ({\bf n} \cdot \widehat{\bf r})$.
If spherical harmonics are used as tensor components, the dipole projection along the $z$ direction is $({\bf n} \cdot \widehat{\bf r})_z = \sqrt{4 \pi/3} \; Y_{{1 0}} \cos \theta$, and the former expression becomes:
\be
= \; \f{4 \pi}{3} R^3 L \sum_{\ell, m} \f{1}{2 \ell + 1} \int Y^*_{\ell m}  d \Omega' \int \left(\sqrt{\tfrac{16 \pi}{5}} \; Y^*_{{2 0}} \; + \sqrt{4 \pi} \; Y^*_{{0 0}} \right) Y_{\ell m} \; d \Omega
\ee
where $\cos^2 \theta$ is written in terms of $Y^*_{2 0}$ and $Y^*_{0 0}$.
This creates two products of Kronecker's deltas, $\delta_{\ell 0} \delta_{m 0}$ and $\delta_{\ell 2} \delta_{m 0}$, only the first of which producing a non-zero contribution:
\be
= \; \f{(4 \pi)^{\f{3}{2}}}{3} R^3 L \sum_{\ell m} \f{\; \delta_{\ell 0} \; \delta_{m 0}}{2 \ell + 1} \int Y^*_{\ell m} \; d \Omega' \; = \; \f{(4 \pi)^2}{3} R^3 L
\ee
The denominator in equation (\ref{eqn:13-b}) follows from equation (\ref{eqn:mode}) with $\ell = 1$:
\be
\iint {\rm G}_i \p_{\bf n} {\rm G}_i d s d s' \; = \; \f{(4 \pi)^2}{3} R
\ee
and the ratio between numerator and denominator is $R^2 L \equiv 2 R^3$, getting:
\be
\lambda^{\rm \; ret}_{\ell = 1} \; = \; \lambda_{\ell = 1} \; - \; 2 \left( \f{k^2}{R} \; + \; {\rm i} k^3 \right) R^3 \; + \; {\rm O} \; (kR)^4
\label{eqn:check}
\ee
i.e. just equation (\ref{eqn:ret}) with the dipole-like eigenvalue, $\lambda_{\ell = 1} = - 2$.\\
\\
iii.) Dipole retardation in a spherical particle may be accounted for from the factor\cite{zeman87} $\kappa = (\tfrac{2}{3} {\rm i} k + 1/R) k^2 \alpha_p$, correcting polarizability by means of radiation damping and depolarization into $\alpha_p / (1 - \kappa)$.
From Clausius \& Mossotti's relation, one has:\cite{jackb}
\be
\f{\alpha_p}{1 - \kappa} \; = \; 
\f{(\epsilon - 1) \; R^3}{\epsilon + 2 - \tfrac{2}{3} {\rm i} (\epsilon - 1) (k R)^3 - (\epsilon - 1) (k R)^2}
\ee
the denominator of which vanishing at:
\be
\epsilon \; (kR) \; = \; - \; \f{2 + \tfrac{2}{3} {\rm i} (k R)^3 + (k R)^2}{1 - \tfrac{2}{3} {\rm i} (k R)^3 - (k R)^2} \; = \; - \; 2 \; - \; 3 (kR)^2 - 2 {\rm i} (k R)^3 \; + \; {\rm O} (kR)^4
\ee
yielding the third-order finite-size corrections to $\epsilon = - 2$.\\
\\
iv.) In equation (\ref{eqn:ret-3}), we need the following average:
\be
I \; \equiv \; \iint X^*_{2 0} Y_{1 0} \; u_2 (1 + u_2) \; X_{2 0} \cos (\theta) \; d \Omega
\ee
of $p_z$ and $d_{z^2}$ terms.
The cubic harmonic $X^*_{2 0} \equiv X_{2 0} = Y_{2 0}$ and, from CFT:
\be
u_2 (R^*) \; = \; \tfrac{7}{3} \sqrt{\pi} \f{\o{r^4}}{{R^*}^4} \left[Y_{4 0} + \sqrt{\tfrac{5}{14}} \left(Y_{4 4} + Y_{4 -4} \right) \right]
\ee
where $R^* = R^* ({\rm S})$ denotes the radius of an equivalent spherical volume.
Finally, the integral returns:
\be
I \; = \; \sqrt{\tfrac{3}{\pi}} \left( \tfrac{2527}{2574} \tfrac{\overline{r^4}}{{R^*}^4} + \tfrac{897}{2574} \right) \tfrac{\overline{r^4}}{{R^*}^4}
\ee
that, in order to recover equation (\ref{eqn:lbo}), should be multiplied by the factor $\sqrt{4 \pi/ 3}$ coming from $({\bf n} \cdot {\bf r})_z$ and the prefactor $- (k R^*)^2/2$. 


\bibliographystyle{rsc}
\bibliography{references}

\end{document}